\documentclass{article}[12pt] 
\usepackage{amssymb} 
\usepackage{amsmath}
\usepackage{amsthm}
\usepackage{amsfonts}
 \usepackage{times}
\usepackage[matrix,arrow]{xy}

\newcommand{\inn}{ ``in'' }
\newcommand{\out}{ ``out'' }
\def\thefootnote{\fnsymbol{footnote}}
\begin{document}
\begin{titlepage}
\today          \hfill
\begin{center}
\hfill hep-th/yymmnnn  \\

\vskip .5in
\renewcommand{\thefootnote}{\fnsymbol{footnote}}
{\Large \bf  Open string radiation from decaying FZZT branes.  }\\
 
\vskip .50in

\vskip .5in
{\large Anatoly Konechny}\footnote{email address: anatolyk@ma.hw.ac.uk}

\vskip 0.5cm
{\large \em Department of Mathematics,\\
Heriot-Watt University,\\
Riccarton, Edinburgh, EH14 4AS, UK \\
and\\
Maxwell Institute for Mathematical Sciences\\
Edinburgh, UK}
\end{center}

\vskip .5in

\begin{abstract} \large
In this paper we continue studying the decay of unstable FZZT branes initiated in 
[1],[2]. The mass of tachyonic mode in this model can be chosen arbitrarily small and 
we use it as a perturbative parameter. In [2] a time-dependent boundary conformal 
field theory (BCFT) describing 
the decay process was studied and it was shown that in a certain sense this BCFT interpolates  
between two stationary BCFT's corresponding to the UV and IR fixed points of the associated 
RG flow. 
In the present work we find in the leading order  vertex  operators 
of the time-dependent BCFT. We  identify the ``in'' and ``out'' vertex 
operators assigned to the UV and IR fixed points and compute the related Bogolyubov coefficients.
We show that there is a codimension one 
subspace of  the out-going states for which pair creation amplitudes are independent of the initial 
wave function of the tachyonic mode. We demonstrate that such amplitudes can be computed within the 
framework of first quantized open string theory via
suitably defined string  two-point functions. 
 We also evaluate a three point function 
which  we interpret as an amplitude for string triplet creation due to interaction. 
 Some peculiarities of scattering amplitudes in the presence of tachyonic modes in the far past are discussed.

\end{abstract}
\end{titlepage}
\large

\newpage
\renewcommand{\thepage}{\arabic{page}}
\setcounter{page}{1}
\setcounter{footnote}{0}
\renewcommand{\thefootnote}{\arabic{footnote}}
\large
\section{Introduction}
\renewcommand{\theequation}{\arabic{section}.\arabic{equation}}

Time dependent backgrounds in string theory are at  present very poorly understood. 
As a first step one may wish to understand perturbative string amplitudes in exact time dependent 
backgrounds. Constructing perturbative amplitudes for such models involves  
 making sense of functional integrals over string worldsheets 
that involve a field with a negative-definite metric. The last one  describes
 a time-like direction in  target space. Assuming that the ghost sector is factorized and that  such integrals are defined by means of a suitable Wick 
 rotation or otherwise, the matter part of an exact background  is described by some non unitary two-dimensional  conformal field theory (CFT).  
One can then consider vertex operators corresponding to infinitesimal deformations of the CFT at hand and 
define string amplitudes in the usual way by integrating the CFT correlators over the  moduli space of 
punctured Riemann surfaces. 

There is next a question of physical interpretation for such amplitudes. Clearly one should seek 
an S-matrix type interpretation. It seems natural to us  to try following in this task 
as closely as possible the analogous field theoretical constructions. It is not our goal 
in this paper to put forward a general string-theoretic scheme for scattering in time-dependent 
backgrounds. Rather we will study in detail one particular model, which is  well-controlled analytically, 
and  
for which we will be able to extend the main field theoretical constructions such as  \inn and \out 
physical states, Klein-Gordon type conserved inner product, Bogolyubov coefficients and particle 
creation amplitudes. For this model we will establish a relation between
 a  string theoretic two point function and a tachyon pair creation amplitude. 
 We will also compute a string  three point function and conjecture its interpretation 
in terms of particle creation amplitudes.

As our considerations  will essentially go in parallel with the field 
theoretical set up it will be instructive to recount it first. This material is fairly  
standard so we will be brief 
(see \cite{BD}, \cite{Wald} for a comprehensive discussion). This will be followed by a short discussion 
of how much of this set up can be brought over into string theory in a straightforward way and 
what are the problems related to the rest of the machinery. 
A disinterested reader may wish to go directly to the next section where the main body of the paper starts.

Consider for definiteness a scalar field $\phi(x)$ with 
a cubic interaction in a non-stationary spacetime. 
To define an $S$-matrix one typically restricts oneself to   
 globally hyperbolic spacetimes and  assumes that the interaction is switched off adiabatically 
for $t\to \pm \infty$. The solution to the equation of motion 
\begin{equation}
[\Box_{x} + m^{2} + \xi R_{x}]\phi(x)  + \lambda \phi^{2}(x) = 0
\end{equation}  
is then assumed to become asymptotically free
\begin{eqnarray}
&& \lim\limits_{t\to -\infty}\phi(x) = \phi_{in}(x) \, , \\
&& \lim\limits_{t\to +\infty}\phi(x) = \phi_{out}(x) 
\end{eqnarray}
where 
\begin{equation} \label{freeKG}
[\Box_{x} + m^{2} + \xi R_{x}]\phi_{in(out)}(x)=0\, .
\end{equation}
To construct the \inn and \out Fock spaces one needs to define which solutions 
 of the free asymptotic equations annihilate the vacuum. In certain spacetimes there 
 are  natural definitions of positive frequency asymptotic solutions which provide a natural definition of  
the \inn and \out vacua. For example if the spacetime at hand is asymptotically 
stationary both in the far past and future,
 i.e. has asymptotic time-like Killing vectors for $t\to \pm \infty$, it is natural to define positive 
 frequency solutions as appropriate eigenstates of those Killing vectors.  
 
 Given a definition of positive frequency modes the construction of the \inn and \out Fock spaces 
 goes as follows.  
  A conserved scalar product on the space of solutions to the  free equation (\ref{freeKG}) is defined as 
 \begin{equation}\label{prod}
 (\phi_{1}, \phi_{2})= -i\int\limits_{\Sigma}(\phi_{1}\partial_{\mu}\phi_{2}^{*} - \phi_{2}^{*}\partial_{\mu}
 \phi_{1})\sqrt{-g}d\Sigma^{\mu}
 \end{equation}  
 where $\Sigma^{\mu}$ is the future directed surface element to a Cauchy surface $\Sigma$. 
 Consider two complete sets of solutions  $u_{p}$, $u_{p}^{*}$ and $v_{q}$, $v_{q}^{*}$ 
 satisfying 
 \begin{equation}\label{oc}
 (u_{p}, u_{p'})=\delta_{p,p'}\, , \quad (u_{p}^{*}, u_{p'}^{*})=-\delta_{p,p'} \, , \quad 
 (u_{p}, u_{p'}^{*})=0
 \end{equation}
and similarly for $v_{q}$. In addition we assume that the modes $u_{p}$ are purely positive frequency 
as $t\to -\infty$ and the modes $v_{q}$ are purely positive frequency as $t\to +\infty$. 
We can expand the \inn and \out fields as 
\begin{eqnarray}
&& \phi_{in}(x)= \sum\limits_{p}[ a^{\rm in}_{p}u_{p}(x) + 
a^{{\rm in}\dagger}_{p}u_{p}^{*}(x)]\, , \nonumber \\
&& \phi_{out}(x)=\sum\limits_{q}[ a^{\rm out}_{q}v_{q}(x) + a^{{\rm out}\dagger}_{q}v_{q}^{*}(x)] \, . 
\end{eqnarray}  
The in and out Fock space vacua are defined by 
\begin{equation}
a^{\rm in}_{p}|0\rangle_{\rm in}=0\, , \qquad a^{\rm out}_{q}|0\rangle_{\rm out}=0
\end{equation}  
and the multiparticle states as 
\begin{eqnarray}
&& |p_{1}\dots p_{n}\rangle_{in} =\prod\limits_{i=1}^{n} 
a^{{\rm in}\dagger}_{p_{i}}|0\rangle_{\rm in} \, , \nonumber \\
&& |q_{1}\dots q_{m}\rangle_{out} =\prod\limits_{i=1}^{m} 
a^{{\rm out}\dagger}_{q_{i}}|0\rangle_{\rm out} \, . 
\end{eqnarray}
The $S$-matrix elements are then defined as overlaps 
\begin{equation}\label{Sgen}
{}_{\rm out}\!\langle q_{1}\dots q_{m}|p_{1}\dots p_{n}\rangle_{\rm in} \, . 
\end{equation} 
In particular the amplitudes of the form 
\begin{equation}
{}_{\rm out}\!\langle q_{1}\dots q_{m}|0\rangle_{\rm in}
\end{equation}
are the particle creation amplitudes.

Since the sets of modes $u_{p}$, $u_{p}^{*}$ and $v_{q}$, $v_{q}^{*}$ are each complete 
they are related by a Bogolyubov transformation:
\begin{eqnarray}\label{Boggen}
&& u_{p}=\sum\limits_{q}( \alpha_{p,q}v_{q} + \beta_{p,q}v_{q}^{*}) \, , \nonumber \\
&& v_{q}= \sum\limits_{p}( \alpha^{*}_{p,q}u_{p} - \beta_{p,q}u_{p}^{*})\, . 
\end{eqnarray}
The orthogonality conditions (\ref{oc}) imply a number of relations between the 
Bogolyubov coefficients $\alpha_{p,q}$, $\beta_{p,q}$:
\begin{eqnarray}\label{st_rels}
&& \sum\limits_{q}(\alpha_{p_{1},q}\alpha^{*}_{p_{2},q} - \beta_{p_{1},q}\beta^{*}_{p_{2},q}) 
=\delta_{p_{1},p_{2}}\, , \quad 
\sum\limits_{q}(\alpha_{p_{1},q}\beta_{p_{2},q} - \beta_{p_{1},q}\alpha_{p_{2},q}) 
= 0\, , \nonumber \\
&& \sum\limits_{p} ( \alpha_{p,q_{1}}^{*}\alpha_{p,q_{2}} - \beta_{p,q_{1}}\beta^{*}_{p,q_{2}}) 
=\delta_{q_{1},q_{2}} \, , \quad 
 \sum\limits_{p} ( \alpha_{p,q_{1}}^{*}\beta_{p,q_{2}} - \alpha^{*}_{p,q_{2}}\beta_{p,q_{1}}) 
= 0\,  .
\end{eqnarray}

One can show that in the case of non-interacting theory ($\lambda=0$) all of the $S$-matrix amplitudes 
(\ref{Sgen}) are expressible via the Bogolyubov coefficients.  Only even numbers 
of particles can be created in this case. 
In particular  pair creation amplitudes can be expressed as 
\begin{equation}
{}_{\rm out}\!\langle q_{1} q_{2}|0\rangle_{\rm in} = 
{}_{\rm out}\!\langle 0|0\rangle_{\rm in}\sum_{p} \beta_{p,q_{1}}^{*} (\alpha^{-1})_{q_{2},p}^{*} \, . 
\end{equation}
Note that relations (\ref{st_rels}) imply that the operator $\alpha_{p,q}$ has a bounded inverse 
$(\alpha^{-1})_{q_{2},p}$ \cite{Ber}. 

In the interacting case one can define a modified 
set of Feynman rules and reduction formulas \cite{BD}, \cite{BT}. In those rules 
one essentially separates the interaction effects  which are taken care of by a suitably defined 
$S$-matrix operator and the effects due to the explicit time-dependence which are encoded in Bogolyubov's 
coefficients. Interaction causes additional particle creation. Thus in $\phi^3$ theory there is 
a  tree level triple creation process (see \cite{BF}, \cite{BT} for some explicit computations).    

Let us remark that although in the above the time dependence was coming from a time-dependent 
space-time metric, most of the discussion generalizes to other instances of time dependence. 
For example one can consider an interacting scalar field in  flat spacetime coupled to 
an external time-dependent potential $V(x,t)$:
\begin{equation} \label{KGpot}
[\Box_{x} + m^{2} + V(x,t)]\phi(x)  + \lambda \phi^{2}(x) = 0 \, . 
\end{equation}
All one needs in order to extend the above discussion to this case is some definition of positive frequency modes. 
For example in a case when the potential vanishes (or goes to a constant) at $t\to \pm \infty$  the definition  
is obvious. 
 Note that the inner product defined in (\ref{prod}) is also conserved on solutions to 
(\ref{KGpot}).

We now turn to string theory. What follows contains some speculations concerning 
 the structure of general formalism of perturbative string theory in time-dependent backgrounds. 
 While supported by  known examples these speculations should be taken as such.   
 We first remark that  natural analogues of wave functions are string physical 
states whose matter part  we denote by $|V\rangle$  which are annihilated by the positive modes of 
Virasoro algebra: 
$L_{n}|V\rangle=\bar L_{n}|V\rangle=0$, $n>0$ and satisfy the mass shell condition
\begin{equation}\label{onshell}
(L_{0} + \bar L_{0})|V\rangle = 2 |V\rangle   \, . 
\end{equation}
The last one is the direct analogue of the free wave equation (\ref{freeKG}).
The $2$ in the right hand side of (\ref{onshell}) should be changed to $1$ for the case of open strings. 
In the simplest situation the zero mode $t$ of the time-like field on the worldsheet 
provides us with a    macroscopic 
time and we can consider the $t\to \pm \infty$ asymptotic regions 
of the target space as the regions where 
we may be able to set up the  \inn and \out scattering states. 
One next would want to specify positive frequency modes in the asymptotic regions. 
Like in the case of quantum field theory \cite{BD}, \cite{Wald}  such definitions 
depend on the particulars of the physical  problem at hand.   One can imagine  
asymptotic time-like Killing vectors to be replaced by  worldsheet currents 
$J^{\alpha}_{\pm}(z,\bar z)$ 
which are  conserved in the asymptotic $t\to \pm \infty$  regions 
 of the string Hilbert space. That is $\partial_{\alpha}J^{\alpha}_{\pm} \sim 0$ for 
 $t\to \pm \infty$. 
  In this case a basis for  incoming positive frequency states 
  can  be defined in terms of  on-shell states which are eigenvectors of  the asymptotic charge 
  $$ \Omega_{-} =  i\lim\limits_{t\to -\infty}\int\!\! dz^{\alpha}(J_{-})_{\alpha} $$
  of the eigenvalue $\omega_{-}$ with $\omega_{-}\ge 0$.
Analogously one defines positive frequency out states as eigenvectors of positive eigenvalue for the asymptotic charge 
$\Omega_{+}$ set up in the $t\to + \infty $ region. 
If $V$ is a vertex operator creating a positive frequency state $|V\rangle$ then its Hermitean conjugate $V^{\dagger}$ 
creates a negative frequency state denoted $|V\rangle^{*}$. 

As in the  case of field theory we can  specify analogs of  one-particle \inn and \out states using a conserved inner product. 
In general such an inner  product (as well as the second quantization symplectic form) 
can be derived from a string field theory kinetic term. The inner product takes a particularly 
simple form when ghosts are factorized and 
 the matter CFT operator $L_{0}$ has the form 
$$
L_{0}=\partial^{2}_{t} + \tilde L_{0}
$$
where $i\partial_{t}$ is the time-like field zero mode momentum operator and  $\tilde L_{0}$ is assumed to be unitary with respect to the BPZ inner product 
$\langle ... \rangle_{\rm BPZ}$  in the CFT state space
\cite{BPZ}. In this situation  one can define a 
conserved hermitean inner product on the space of solutions to the on-shell condition  (\ref{onshell}) as 
\begin{equation} \label{inpr_str}
( V_{1}, V_{2}) \equiv \frac{i}{2}[\langle V_{1}|\partial_{t} V_{2}\rangle_{\rm BPZ} -  
\langle V_{2}|\partial_{t} V_{1}\rangle_{\rm BPZ}]\,  .
\end{equation}  

We can now pick  bases of positive frequency states 
$| P\rangle_{\rm in}$, $| Q\rangle_{\rm out}  $ that together with the conjugate states 
$| P\rangle_{\rm in}^{*}$, $| Q\rangle_{\rm out}^{*}  $ satisfy conditions similar to (\ref{oc}) 
with respect to the inner product 
(\ref{inpr_str}). Here $P$ and $Q$ are 
complete sets of asymptotic quantum numbers labeling the positive frequency \inn and \out states 
respectively. 
The negative frequency states  $| P\rangle_{\rm in}^{*}$, $| Q\rangle_{\rm out}^{*}  $ 
 can be interpreted as incoming or respectively  outgoing string fundamental excitations or 
 particles. Let us denote by $V_{P}^{\rm in}$ and $V_{Q}^{\rm out}$ the  world sheet 
 vertex operators corresponding to the states $| P\rangle_{\rm in}$, $| Q\rangle_{\rm out}  $. 
 Assuming both bases are  complete these operators are related by  Bogolyubov 
 transformations 
 \begin{eqnarray}
&& V_{P}^{\rm in}=\sum\limits_{Q}( \alpha_{P,Q}V_{Q}^{\rm out} + 
\beta_{P,Q}V_{Q}^{\rm out \dagger}) \, , \nonumber \\
&& V_{Q}^{\rm out}= \sum\limits_{P}( \alpha^{*}_{P,Q}V_{P}^{\rm in} - 
\beta_{P,Q}V_{P}^{\rm in \dagger})\, . 
\end{eqnarray}
Define operators 
$$
{\cal V}_{Q}^{\rm in} = \sum\limits_{P}(\alpha^{-1})_{Q,P}V_{P}^{\rm in} \, . 
$$
These operators are pure positive frequency in the far past and in the far future 
their positive frequency part is $V_{Q}^{\rm out}$. 
It seems natural to us to conjecture that 
the string theoretic two-point function of such operators gives a  normalized pair creation amplitude
\begin{equation}\label{2ptconj}
\frac{1}{2}\langle {\cal V}_{Q_{1}}^{\rm in} {\cal V}_{Q_{2}}^{\rm in} \rangle_{\rm str}= 
\sum\limits_{P} \beta_{P,Q_{1}} (\alpha^{-1})_{Q_{2},P} = 
\frac{{}_{\rm in}\!\langle 0|Q_{1} Q_{2}\rangle_{\rm out}}{  
{}_{\rm in}\!\langle 0|0\rangle_{\rm out}} \, . 
\end{equation}
A similar conjecture was put forward  in \cite{GS} regarding certain CFT two-point functions and pair creation rates. 
While it may work for a certain type of models such as time-like Liouville Theory considered in  \cite{GS}, it 
seems to us that in general, when the time-like part of the CFT and the spatial part are mixed 
( as in  the model considered in this paper) one should consider the string two-point function.  
The main difficulty with this proposal is 
technical.  There is no general prescription for computing a string-theoretic two-point function. 
 A formal expression for a two-point function in a noncompact target space contains 
 an infinity coming from integrating over target space zero modes and a 
 vanishing factor arising from the division by  the infinite volume of the group of worldsheet modular transformations 
  fixing two  points. The cancellation of the two infinities is relatively well understood only for noncritical strings \cite{Seiberg}, \cite{GM}, \cite{Kostov}, 
 \cite{DFrKut}. 
 
 String theory is an interacting theory. There should be prescriptions to compute multiparticle S-matrix amplitudes of the type (\ref{Sgen}) that 
 take into account string interactions. We conjecture that whenever string $n$-point functions of operators ${\cal V}_{Q}^{\rm in}$ can 
 be defined they give perturbative contributions (at each genus) to the  $n$-particle creation amplitudes 
 $$
 \langle {\cal V}_{Q_{1}}^{\rm in} {\cal V}_{Q_{2}}^{\rm in} \dots  {\cal V}_{Q_{n}}^{\rm in} \rangle_{\rm str} = 
C_{n} \frac{{}_{\rm in}\!\langle 0|Q_{1} Q_{2}\dots Q_{n}\rangle_{\rm out}}{  
{}_{\rm in}\!\langle 0|0\rangle_{\rm out}} 
$$
where $C_{n}$ is a numerical normalization factor.

In the above discussion there were no specific points referring to closed strings so if correct the same scheme should 
apply also to open string time-dependent backgrounds. Also very mild assumptions were made on the nature of the time-dependence. 
In this paper we consider a particular model describing a time dependent process of open string tachyon condensation in two-dimensional string theory. 
We will find that the presence of 
tachyon instability in the initial system brings about certain additional complications into the above general scheme. Namely 
solutions exponentially growing in the far past are needed for completeness of the  out scattering states. We suspect this  
to be a generic situation for  decays of unstable backgrounds. This results on the one hand  in  an additional ambiguity in defining 
the initial state of the system and on the other hand, from the CFT perspective, in the need to define correlation functions for 
exponentially blowing up operators. Nevertheless we will show that for a large class of vertex operators (for a codimension one subspace in the total space 
of solutions) the correlation functions do not depend on these additional ambiguities. We will  demonstrate that 
for these  states  the appropriately defined string two point functions give   pair creation amplitudes. 
We will also compute  a string three point amplitude  and  conjecture its interpretation in terms of particle triplet creation rate.    
A more detailed discussion of our results is given in the final section of 
the paper. 

The main body of the paper is organized as follows. In section 2 we introduce the model we study and give a review of main results obtained 
in \cite{1}, \cite{2}. In section 3 we explain our approach to constructing vertex operators in our time-dependent model and compute explicitly 
vertex operators asymptoting to plane waves in the infinite  past. In section 4 we discuss the conserved inner product and define normalized 
\inn and \out states. In section 5 a basis of vertex operators asymptoting to positive frequency out states are constructed  and the exponential blow 
up of such solutions in the infinite past is demonstrated. In section 6 we analyze the Bogolyubov transformation relating the \inn and \out vertex 
operators. In section 7 we develop the secondary quantization of the model. We construct a family of 
physically natural initial states in the oscillators state space. 
In section 8 a string theoretic two point function is computed in a certain regularization scheme. It is shown that for a large (codimension one) 
class of out states it gives a pair production rate. In section 9 a string three point function is computed. In the final section we discuss 
our results and point at some open questions. The appendix contains some technical details of the computations.  


\section{The model}\label{model}
\setcounter{equation}{0}
In this section we explain the particulars of the model 
and review the main results obtained in \cite{1}, \cite{2}. 
Our model is constructed in the framework of  $c=1$ noncritical string theory 
(see e.g. \cite{GM} for a review). The worldsheet 
CFT is a product of a free timelike boson $X_{0}$ and the $c=25$ Liouville theory. 
The action for $X_{0}$ is
\begin{equation}
S_{X_{0}}= -\frac{1}{4\pi}\int d^2 x\, (\nabla X_{0})^{2} 
\end{equation} 
where the sign in front of the action means that $X_{0}$ is timelike. 
The Liouville theory is a conformal field 
theory of an interacting noncompact boson $\phi$ with the action functional
\begin{equation}
S_{L}=\frac{1}{4\pi}\int d^{2} x [(\nabla \phi)^{2} + 4\pi \mu e^{2b\phi}]
\end{equation}
and the background charge $Q=b + 1/b$ is introduced via the the asymptotic at spatial 
infinity $\phi(x) \sim -Q\log x^{2}$.   
This theory  is by now well understood 
and we refer the reader to \cite{TeschnerL} for a review of essential results.
From now on we set $b=1$ that corresponds to the central charge $c=25$.
At the level of sigma model description the $c=25$ Liouville theory is characterized 
by a flat 2D metric and the following backgrounds for the dilaton and tachyon fields $\Phi$ and $T$: 
\begin{equation}
\Phi(\phi, X_{0})  = \phi \, , \qquad T(\phi, X_{0})=\mu e^{2\phi} \, . 
\end{equation}
The linear dilaton profile implies that the string coupling goes to zero in the $\phi \to -\infty$ 
region and blows up in the $\phi \to + \infty$ region. The strings however are essentially confined 
to the weakly coupled region by the tachyon potential. String scattering in and out states are 
naturally set up in  the $\phi \to -\infty$  asymptotic region.

We are further interested in the open string sector of this theory which is introduced via 
Neumann type conformal boundary conditions. 
The corresponding D1-branes are referred to in the 
literature as FZZT branes after the authors of \cite{FZZ}, \cite{T}.  At the semiclassical level 
these boundary conditions are defined by adding to the bulk theory put  on an upper half plane 
$\{(x,\tau)|x\in {\mathbb R}, \tau \ge 0\}$ a boundary 
action  
\begin{equation} \label{boundaryL}
S_{\partial} = \mu_{B} \int_{\mathbb R} dx\, e^{\phi} 
\end{equation} 
 which results in the boundary conditions
 \begin{equation}
 i(\partial - \bar \partial)\phi = 2\pi \mu_{B} e^{\phi} \, . 
 \end{equation}
By solving the corresponding boundary conformal field theory (BCFT) 
exactly it was found that at the quantum level 
for each value of the boundary coupling $\mu_{B}$ there are countably many physically 
distinct boundary conditions \cite{FZZ}, \cite{T}, \cite{PT}, \cite{1}. The quantum 
boundary conditions are parameterized by a parameter $\delta$
 related to $\mu_{B}$ via\footnote{The parameter $\delta$ is 
related to the parameter $\sigma$ from \cite{PT}, \cite{1} as $2\sigma=1-\delta$ and to the parameter 
$s$ from \cite{FZZ} as $s=i(1 + \delta)$.} 
\begin{equation}
\cos[\pi(1 + \delta)]= \frac{\mu_{B}}{\sqrt{\mu}} \, . 
\end{equation}



In particular the spectrum of boundary operators depends on the value of $\delta$. 
Before we discuss the latter a note on the choice of worldsheet is in order. 
For the most part of the paper 
our computations will be done on a disc 
${\mathbb D}=\{(r, \sigma)| 0\le r\le 1, 0\le \sigma \le 2\pi   \}$. This 
 concerns in particular the computations of the two and three point functions in sections 
 \ref{2ptsec} and \ref{3ptsec}.  To use the state-operator correspondence in a BCFT we invoke  
 a Hamiltonian quantization on a strip ${\mathbb S}=\{ (\sigma, \tau)| 0\le \sigma \le \pi \}$ 
 with the boundary condition specified by $\delta$ imposed on both edges of the strip. 
 The corresponding Hilbert space has the form 
 \begin{equation} \label{spectrum}
 {\cal H}_{\delta \delta}^{B} = \int\limits_{{\mathbb R}_{+}}\, dP\, \,  {\cal V}_{P} \oplus 
 \left\{ \begin{array}{lll}
 \emptyset & \mbox{for } &\delta<0 \, , \\
 {\cal V}_{\vartheta} & \mbox{for } &\delta >0 \, , 
 \end{array} \right. 
 \end{equation}
 where $\vartheta = i\delta$ and ${\cal V}_{Q}$ is the irreducible unitary 
 representation of the Virasoro algebra 
 with $c=25$ and the highest weight $\Delta_{Q} = 1 + Q^{2}$. 
 The above representation means that for 
 each value of $\delta$ there is a continuum of $\delta$-function normalizable 
states with weights $\Delta_{P}=1 + P^{2}\, ,P\in {\mathbb R}_{+}  
$ whose boundary fields we denote $\Phi_{P}^{\delta}(x)$. 
   In addition for 
$\delta >0$ there is a single discrete normalizable state whose vertex operator we 
denote by $\Phi_{\vartheta}^{\delta}(x)$. The conformal weight of this operator 
is $\Delta_{\vartheta}=1-\delta^{2}$. In many manipulations it can be treated 
on equal footing with the operators $\Phi_{P}^{\delta}$   if one regards it as 
an operator $\Phi_{P}^{\delta}$ with $P=\vartheta=i\delta$. 
We choose normalizations of our operators as in \cite{2} to be given in more detail shortly.

For the $X_{0}$ field we consider the Neumann boundary condition. 
The corresponding Virasoro primary states   are  denoted $|\omega\rangle_{X_{0}}$:
\begin{equation}
L_{0}^{open}|\omega\rangle_{X_{0}} = -\omega^{2}|\omega\rangle_{X_{0}}\, \quad 
L_{n}^{open}|\omega\rangle_{X_{0}} =0\, , \enspace n>1\, . 
\end{equation}
We introduce  zero modes of the fields as 
\begin{equation} 
t=\int_{0}^{\pi}d\sigma \, X_{0}(\sigma, 0)\,  \quad 
\phi_{0}^{op}=\int_{0}^{\pi}d\sigma\, \phi(\sigma, 0) \, .
\end{equation} 
This allows us to define the wave functions for the highest weight 
states $|P\rangle \otimes |\omega \rangle_{X_{0}} $  as 
\begin{equation} \label{wfns}
\Psi_{P}(\phi_{0}^{op},t)= \langle \phi_{0}^{op} |P\rangle \cdot \langle t |\omega \rangle_{X_{0}} = 
\langle \phi_{0}^{op} |P\rangle \cdot e^{i\omega t}
\end{equation}
 and similarly with $P$ replaced by $\vartheta$. 
 In the weak coupling region $\phi_{0}^{op}\to -\infty$ the Liouville parts of wavefunctions 
 behave as 
 \begin{equation}\label{Psi_asympt}
 \Psi_{P}(\phi_{0}^{op}) \sim C_{\delta}(P)e^{iP\phi_{0}^{op}} + C_{\delta}(-P)e^{-iP\phi_{0}^{op}}
 \end{equation}
  where $C_{\delta}(P)$ is a certain normalization factor. One can show (see e.g. \cite{2} Appendix B.1) 
  that $C_{\delta}(\vartheta)=0$ and thus the wavefunction $\Psi_{\vartheta}(\phi_{0}^{op})$ 
  decays exponentially in the asymptotic region that is characteristic of a bound state.
  
Since $\Delta_{\vartheta}$ is less than one the corresponding open string state is tachyonic. 
The string spectrum contains an unstable excitation with the vertex operator 
$\Phi^{\delta}_{\vartheta} e^{\delta X_{0}}$ 
whose wavefunction increases exponentially with $t$. We thus have a system with a localized open string 
tachyon whose mass $\delta$ can be chosen to be arbitrarily small. From the target space perspective 
the smallness of  $\delta$  means that the tachyon condensation process is slow rolling. This process 
can be described by  deforming the $({\rm Liouville})\times X_{0}$ BCFT  adding to it a boundary 
interaction term generated by the tachyon vertex operator
\begin{equation} \label{interaction}
S_{\lambda}=\lambda \int_{\mathbb R}dx \, [\Phi^{\delta}_{\vartheta} e^{\delta X_{0}}](x) \, . 
\end{equation}  
Looking at the operator product expansions of the multiple products of the tachyon operator with itself 
it is easy to see that no divergences arise when treating the interaction term perturbatively. The deformed 
theory is therefore conformal. The smallness of the parameter $\delta$ can then be used by employing the 
RG resummation technique to construct an effective Lagrangian that gives the boundary state to the leading 
order in $\delta$ \cite{2}. 

For future reference we give 
here the details of the spectrum of boundary operators of the FZZT branes in the $\delta \to 0$ limit. 
The two point functions in the normalizations of \cite{2}
are 
\begin{equation}\label{2p1}
\langle \Phi^{\delta}_{P}(x_1) \Phi^{\delta}_{P'}(x_2)\rangle = 
|x_1 - x_2|^{-2\Delta_{P}} C_{\delta}(P)C_{\delta}(-P)\delta(P-P')   \, , 
\end{equation}
\begin{equation}\label{2p2}
\langle \Phi^{\delta}_{\vartheta}(x_1) \Phi^{\delta}_{\vartheta}(x_2)\rangle = 
|x_1 - x_2|^{-2\Delta_{\vartheta}} d_{\delta}
\end{equation}
 where the factors $C_{\delta}(P)$ and $d_{\delta}$ in the leading order are   
 \begin{equation} \label{Cd}
 C_{\delta}(P) \sim \mu_{r}^{1/2}\frac{\pi(\delta + iP)}{iP} \, , \quad 
 d_{\delta} \sim \frac{\pi \mu_{r}}{\delta} \, . 
 \end{equation}
 Here and elsewhere when taking the $\delta \to 0$ asymptotics appropriate for conformal perturbation theory 
 one should assume that the Liouville momenta $P$ are all of the order $\delta$ (see \cite{2} for a detailed 
 explanation). We will be often using the rescaled variables 
 \begin{equation}
 p=P/\delta \, \quad q=Q/\delta \, , \enspace \mbox{etc.}
 \end{equation}
 which are of the order $\delta^{0}$. 
 
 The operator product expansions have the form
 \begin{eqnarray}
 \Phi_{P_2}^{\delta}(x_2)\Phi_{P_1}^{\delta}(x_1)=&& \int\limits_{0}^{\infty}dP_{3}\, F_{P_{2}P_{1}}^{P_{3}}
 |x_2 - x_1|^{\Delta_{P_{3}}-\Delta_{P_{2}} -\Delta_{P_{1}}} \Phi_{P_{3}}^{\delta}(x_1) + 
 \nonumber \\
 && f_{P_{2}P_{1}}^{\vartheta}|x_{2}-x_{1}|^{\Delta_{\vartheta}-\Delta_{P_{2}} -\Delta_{P_{1}}}
 \Phi_{\vartheta}^{\delta}(x_1) + \enspace \mbox{descendants} 
 \end{eqnarray}
 where $P_{2}$ and $P_{1}$ can also assume the value $\vartheta$.
 The asymptotic formulas for the OPE coefficients are 
 \begin{equation} \label{OPE1}
 F_{P_{2}P_{1}}^{P_{3}} \sim \frac{2P_{3}^{2}}{\pi(\delta^{2} + P_{3}^{2})} \, , 
 \end{equation}
 \begin{equation}\label{OPE2}
 f_{P_{2}P_{1}}^{\vartheta} = -2\pi i {\rm Res}_{P3 = i\delta}F_{P_{2}P_{1}}^{P_{3}} \sim 2\delta 
 \end{equation}
 where $P_{1}$ and $P_{2}$ can take the value $\vartheta$. 
 
 The three point functions to the leading order in $\delta$ all take the same value 
 \begin{equation}\label{3ptL}
 \langle \Phi_{P_1}^{\delta}(x_1) \Phi_{P_2}^{\delta}(x_2) \Phi_{P_3}^{\delta}(x_3)\rangle \sim 2\pi 
 |x_1 - x_2|^{\Delta_{12}} |x_2 - x_3|^{\Delta_{23}} | x_3 - x_1|^{\Delta_{13}}  
 \end{equation}  
 where $\Delta_{ij}$ are  standard combinations of conformal dimensions. 
 In the last formula any of $P_{i}$'s can take the value $\vartheta$.

 The leading order contributions from  interaction (\ref{interaction}) to correlation functions 
 come from short distances. Hence, although the theory is finite, one can use the RG resummation technique 
 to construct the effective Lagrangian \cite{2}. Introducing a short distance cutoff $\epsilon$ we 
 write down a renormalized boundary action  that includes all operators near-marginal in the $\delta \ll 1$ 
 limit which are generated via short distance expansions
 \begin{equation}\label{Sren_lambda}
 S_{\rm \lambda}^{\rm ren}=\sum\limits_{n=1}^{\infty}\int\limits_{\mathbb R}\! dx\, 
 \Bigl( U_{n}\epsilon^{(n^2-1)\delta^{2}}
 [e^{n\delta X_{0}}\Phi_{\vartheta}^{\delta}](x) + \int\limits_{0}^{\infty}dP\, \lambda_{n}(P)
 \epsilon^{n^{2}\delta^{2} + P^{2}}[e^{n\delta X_{0}}\Phi_{P}^{\delta}](x) \Bigr). 
 \end{equation}    
 The RG equations arise as conditions for $\epsilon$-independence of the correlation functions. 
 These equations supplemented by the appropriate  conditions fixing the bare couplings 
 can be solved explicitly \cite{2}. The results are most elegantly summarized in terms 
 of generating functions 
 \begin{equation} \label{genfun1}
 \lambda(q,t) = \sum\limits_{n=1}^{\infty}\lambda_{n}(q\delta)e^{n\delta t} \, , \quad 
 U(t)=\sum\limits_{n=1}^{\infty} U_{n}e^{n\delta t} 
 \end{equation}
 where the parameter $t$ can be identified with the zero mode of the $X_{0}$ field. 
 The function $U(t)$ can  be expressed via elementary functions and the function 
 $\lambda(q,t)$ via the hypergeometric function ${}_{2}F_{1}$. 
 The explicit expressions can be found in \cite{2} and will not be used in this paper.

 A combination of the generating functions that will be important later is
 \begin{equation}\label{Wdef}
 W(t)=U(t) + \delta \int\limits_{0}^{\infty}dq\, \lambda(q, t) \, .
 \end{equation}
It has a simple explicit expression 
 \begin{equation}\label{W}
 W(t) =  \delta \Bigl( \frac{\nu e^{\delta t}}{1 + \nu e^{\delta t}} \Bigr)
 \end{equation}
 where $\nu =\lambda/\delta$ - the rescaled bare coupling $\lambda$ from (\ref{interaction}). 
 
 Although the generating functions  (\ref{genfun1}) ab initio  had a radius 
 of convergence bounded by $t \ll \delta^{-1}$ they have a natural analytic 
 continuation for all values\footnote{It is interesting to note in regard with the analytic continuation that 
 it is well defined only for $\nu>0$. For negative values of $\nu$ the analytically continued solution will hit a branch cut for 
 sufficiently large values of $t$. This can be correlated with the fact that there is no perturbative fixed point for the RG flow 
 triggered by the relevant  operator $\Phi_{\vartheta}^{\delta}$ with a negative coupling constant} of $t$. In particular one can find the 
 $t\to \infty$ asymptotic which can be interpreted as the far future of the 
 FZZT brane decay process. The generating function $U(t)$ tends 
 to a constant $u_{*}=2\delta$ while the couplings of the continuous 
 operators have asymptotics 
 \begin{equation} \label{lambda_as}
 \lambda(q,t) \to -\frac{2}{\pi(1 + q^{2})} + \frac{q}{\sinh(\pi q)} 
 \Bigl( \frac{e^{iq\delta t}\nu^{iq}}{1 + iq} + 
 \frac{e^{-iq\delta t}\nu^{-iq}}{1 -iq} \Bigr)\, .  
 \end{equation}
 It was further shown that the constant ($t$-independent) parts of 
 the above asymptotics are described by  the boundary condition 
 characterized by the parameter $-\delta$ while the oscillatory piece in    
 (\ref{lambda_as}) was interpreted as  radiation. The conclusion of 
 \cite{2} was that the $0<\delta \ll 1$ FZZT brane decays into the 
 $\delta_{*}=-\delta$ brane leaving behind a radiation travelling towards   
 $\phi_{0}^{op} = -\infty$.  
 
 It is one of the purposes of the present paper to derive the radiation 
 produced in the decay process from first principles.


\section{Time-dependent vertex operators}\label{vops}
\setcounter{equation}{0}
\subsection{First order conformal deformation equations}

We would like to find marginal operators of the time-dependent BCFT (\ref{interaction}). 
In general marginal operators of a given (B)CFT can be identified with its infinitesimal deformations.   
Consider an infinitesimal perturbation of 
the boundary theory (\ref{boundaryL}), (\ref{interaction}) by a term
$$
\xi_{0} \int_{\mathbb R} dx\, [e^{ iPX_{0}}\Phi^{\delta}_{|P|}](x)  
$$ 
where $\xi_{0}$ is a constant (the  deformation parameter) 
and $P$ is any real number except zero. The $P=0$ case needs special care and will 
be treated separately later. 
The perturbing operator $e^{ iPX_{0}}\Phi^{\delta}_{|P|}$ is a primary of  dimension $1$ relative 
to the undeformed $({\rm FZZT})\times X_{0}$ boundary theory. The tachyon interaction term 
(\ref{interaction}) will result in mixing of this operator with operators  
 $e^{(n\delta +iP)X_{0}}\Phi_{Q}^{\delta}$, $e^{(n\delta + iP)X_{0}}\Phi^{\delta}_{\vartheta}$, 
 $n\in {\mathbb Z}_{+}$ that can be found via operator product expansion. Thus we are  led 
 to consider a combined renormalized action  
  \begin{equation} \label{combined_act}
S^{\rm ren}_{\rm Bd}=S^{\rm ren}_{\lambda} + \Delta S^{\rm ren}_{\mu, \eta}
\end{equation}
where $S^{ren}_{\lambda}$ is given by (\ref{Sren_lambda}) and 
\begin{eqnarray} \label{ren_action2}
&&\Delta S^{ren}_{\mu, \eta} = \int\limits_{\mathbb R}\! dx \, \sum_{n=0}^{\infty} \int dQ\, \mu_{n}(P,Q)
\epsilon^{(n\delta  +iP)^{2}+Q^{2}}[e^{(n\delta +iP)X_{0}}\Phi_{Q}^{\delta}](x) +  \nonumber \\
&&\int\limits_{\mathbb R}\! dx \, \sum_{n=1}^{\infty} \, \eta_{n}(P)\epsilon^{(n\delta + iP)^{2}-\delta^{2}}
[e^{(n\delta  iP)X_{0}}\Phi^{\delta}_{\vartheta}] (x)\, . 
\end{eqnarray}
The last perturbation is considered only to first order in $\mu_{n}(P,Q)$ and $\eta_{n}(P)$.

The beta functions for the couplings $\mu_{n}(P,Q)$ and $\eta_{n}(P)$ can be computed via conformal 
perturbation theory as in \cite{2}. It is clear from the general form of the OPE's at hand that
the RG equations for the original couplings $U_{n}$ and $\lambda_{n}(P)$ are unaffected by the presence of 
the new couplings\footnote{This is true only for  $P\ne 0$.} 
 in    $\Delta S^{\rm ren}_{\mu, \eta}$ and are given by the solution found in \cite{2}.
The additional RG equations for the new couplings read 
\begin{eqnarray} \label{addRG}
&&\epsilon \frac{d \mu_{n}(P,Q)}{d\epsilon} = -(Q^{2}+ (n\delta + iP)^{2})\mu_{n}(P,Q) 
-2\sum_{l=1}^{n} \int\!\! dQ' \, F^{Q}_{\vartheta Q'} U_{l}\mu_{n-l}(P,Q') -  \nonumber \\
&& 2\sum_{l=1}^{n}\Bigl[ \int\!\! \int\!\! dQ' dQ''\,  F^{Q}_{Q'Q''}\lambda_{l}(Q'')\mu_{n-l}(P,Q') +
F^{Q}_{\vartheta P}U_{l}\eta_{n-l}(P) + \int\!\! dQ' \, \lambda_{l}(Q')\eta_{n-l}(P)\Bigr] 
 \nonumber \\
&& \epsilon \frac{d \eta_{n}(P,Q)}{d\epsilon} = (\delta^{2}-(n\delta + iP)^{2})\eta_{n}(P)
 -2\sum_{l=1}^{n-1}f^{\vartheta}_{\vartheta\vartheta}U_{l}\eta_{n-l}(P)- \sum_{l=1}^{n}\Bigl[ 
 \int\!\! dQ\,  f^{\vartheta}_{\vartheta Q} \lambda_{l}(Q) \eta_{n-l}(P) \nonumber \\
&& 
 + \int\!\! dQ\,  f^{\vartheta}_{\vartheta Q}U_{l}\mu_{n-l}(P,Q) + 
 \int\!\!\int\!\! dQ dQ'\,  f^{\vartheta}_{QQ'}\lambda_{l}(Q')\mu_{n-l}(P,Q) \Bigr]\, . 
\end{eqnarray}

The supplementary renormalization conditions are 
\begin{eqnarray}\label{suppRG}
\lim\limits_{\epsilon \to 0} \eta_{n}(P)\epsilon^{(n\delta + iP)^{2}-\delta^{2}} = 0 \, , && \nonumber \\
\lim\limits_{\epsilon \to 0}  \mu_{n}(P,Q)
\epsilon^{(n\delta + iP)^{2} +Q^{2}} =0
\quad \mbox{for} \enspace n>0 \, , && \nonumber \\
\lim\limits_{\epsilon \to 0}  \mu_{0}(P,Q)
\epsilon^{Q^{2}-P^{2}} = \xi_{0} \delta(P-Q) && 
\end{eqnarray}
where $\xi_{0}$ is  constant. 
The above equations with these conditions can be solved 
recursively and are equivalent to 
\begin{equation}
\epsilon\frac{d \mu_{n}(P,Q)}{d\epsilon}=0=\epsilon\frac{d \eta_{n}(P)}{d\epsilon}
\end{equation}
which means, as  anticipated, that the corresponding perturbation generated by a nonvanishing $\xi_{0}$ 
is first order marginal.  

An additional  comment  perhaps is in order on the meaning of the renormalized action (\ref{combined_act}).  
The renormalized action (\ref{Sren_lambda}) is an effective action that can be used to compute 
the leading orders in $\delta$ of the disc partition function and one-point functions of the bulk operators 
of the time dependent BCFT (\ref{interaction}). 
 The renormalized 
action ( \ref{combined_act}) with the constants $\mu_{n}$ and $\eta_{n}$ treated to the first order 
can be used to compute the leading order in $\delta$ of the correlators of marginal {\it boundary} operators 
of the BCFT (\ref{interaction}).  
Solutions to the (\ref{addRG}) substituted into (\ref{ren_action2}) thus give renormalized boundary 
marginal operators labeled by $P$: 
\begin{equation}\label{newops}
\Phi_{P}(\nu)=\sum\limits_{n=0}^{\infty}\int\limits_{0}^{\infty} dQ\, \mu_{n}(P,Q)
[e^{(n\delta +iP)X_{0}}\Phi_{Q}^{\delta}] + 
\sum_{n=1}^{\infty} \, \eta_{n}(P)
[e^{(n\delta  iP)X_{0}}\Phi^{\delta}_{\vartheta}] 
\end{equation}
where we included the coupling constant $\nu$ in the notation to signify  that these are primaries of 
the deformed theory (\ref{interaction}).

In the leading order in $\delta$ one uses the asymptotic expressions for the OPE coefficients 
(\ref{OPE1}), (\ref{OPE2}) in 
the above equations to obtain  
\begin{eqnarray}\label{RGadd}
&&(Q^{2}+ (n\delta + iP)^{2})\mu_{n}(P,Q)=-2f(Q)\sum_{l=1}^{n}W_{l}h_{n-l}(P)\, , \nonumber \\
&&(\delta^{2}-(n\delta + iP)^{2})\eta_{n}(P)=4\delta\sum_{l=1}^{n}W_{l}h_{n-l}(P) 
\end{eqnarray}
where
\begin{equation}
h_{n}(P) =\eta_{n}(P) + \int\limits_{0}^{\infty} dQ\, \mu_{n}(Q,P)\, ,
\end{equation}
\begin{equation}\label{f} 
f(Q)=\frac{2Q^{2}}{\pi(\delta^{2} + Q^{2})}\, 
\end{equation}
and the coefficients $W_{n}$  can be read off the generating function $W(t)$ given in (\ref{Wdef}), (\ref{W}).
Note that while a particular renormalization scheme was used to obtain (\ref{addRG}) the leading order equations 
(\ref{RGadd}) are scheme independent.

The above equations look most compact when written in terms of generating functions
 \begin{eqnarray} \label{genf}
\mu(t,P,Q)&=&\sum_{n=0}^{\infty}\mu_{n}(P,Q)e^{(n\delta + iP)t}\, ,  \nonumber \\
 \eta(t,P)&=&\sum_{n=1}^{\infty}\eta_{n}(P)e^{(n\delta + iP)t}  \nonumber \\ 
 h(t,P)&=&\sum_{n=0}^{\infty}h_{n}(P)e^{(n\delta + iP)t} =\eta(t,P) + \int\limits_{0}^{\infty} dQ\, \mu(t,P,Q)\, . 
 \end{eqnarray}
Identifying as before the parameter $t$ with the target space time (the zero mode of $X_{0}$) 
we can think of these generating functions as time-dependent coupling constants. 
Equations (\ref{RGadd}) read
\begin{eqnarray} \label{diffeqs}
-(Q^{2} + \partial_{t}^{2})\mu(t,P,Q) &=& 2f(Q)W(t)h(t,P) \, , \nonumber \\
(\partial_{t}^{2} -\delta^{2})\eta(t,P)&=&-4\delta W(t) h(t,P) \, .
\end{eqnarray}

Assuming that for $t$ sufficiently small: $t\ll \delta^{-1}$ the series expansions (\ref{genf}) converge 
the renormalization conditions (\ref{suppRG})  imply the initial conditions 
\begin{eqnarray}\label{ics}
\lim_{t\to - \infty} \mu(t,P,Q)&=& e^{ iPt}\xi_{0}\delta(P-Q)  \, \nonumber \\
\lim\limits_{t\to -\infty}\eta^{\pm}(t,P)& =&0\, 
\end{eqnarray}
that means that the wave functions for operators (\ref{newops}) in the infinite past look like 
$$
\xi_{0}e^{i Pt}\Psi_{|P|}(\phi_{0}^{op}) \, .
$$
The conditions (\ref{ics}) are to be used as initial conditions for solving the differential 
equations (\ref{diffeqs}).

Note that unlike the mode equations (\ref{RGadd}) the continuous equations (\ref{diffeqs}) do not carry any reference to the initial condition 
and thus should be regarded as more fundamental. These equations are the direct analogue of the on-shell condition (\ref{onshell}).
Note also their similarity  to (\ref{KGpot}).


\subsection{Solutions asymptoting to plane waves in the far past}
We would like  to find an explicit solution to equations (\ref{RGadd}) with boundary conditions (\ref{suppRG}) or equivalently (\ref{diffeqs}), (\ref{ics}). 
We will be using both forms of equations interchangeably.

With the given boundary conditions (\ref{suppRG}) we can rewrite the first equation in (\ref{RGadd})  as
\begin{equation}\label{RG1}
\mu_{n}(P,Q)=-\frac{2f(Q)}{(Q^{2}+ (n\delta + iP)^{2})}\sum_{l=1}^{n}W_{l}h_{n-l}(P)
+ \xi_{0}\delta(P-Q)\delta_{n,0} \, .
\end{equation}

Introducing
\begin{equation}
\xi_{n}(P)\equiv \int\limits_{0}^{\infty} dQ\, \mu_{n}(P,Q)
\end{equation}
we obtain by integrating (\ref{RG1})   over $Q$
\begin{equation}
(\delta(n+1) + iP)\xi_{n}(P)=-2\sum_{l=1}^{n}W_{l}h_{n-l}(P)  + \delta_{n,0}\xi_{0}(\delta 
+ iP)\, .
\end{equation}

The second equation in (\ref{RGadd}) can be rewritten as
\begin{equation}\label{RGsecond}
(\delta(n+1)+iP)(\delta(n-1) + iP)\eta_{n}(P)=-4\delta\sum_{l=1}^{n}W_{l}h_{n-l}(P) 
\end{equation}
Taking an appropriate linear combination of the last two equations we obtain
\begin{equation} \label{heq}
(\delta(n-1)+iP)h_{n}(P)=-2\sum_{l=1}^{n}W_{l}h_{n-l}(P) + \xi_{0}(-\delta + iP)\delta_{n,0}
\end{equation}
that is an equation on the modes $h_{n}(P)$.  We proceed by solving first this equation and  then substituting  
the solution into (\ref{RG1}), (\ref{RGsecond}) and their continuous counterparts (\ref{diffeqs}).

Switching to the generating functions we  rewrite (\ref{heq}) as
\begin{equation}
(\partial_{t} - \delta )h(t,P)= - 2W(t)h(t,P) + \xi_{0}(-\delta + iP)e^{iPt} \, .
\end{equation}
It is straightforward to find the solution to this equation 
\begin{equation}\label{hsol}
h(t,P)=\xi_{0}\Bigl( \frac{2W^{2}(t)}{\delta^{2}ip(1 +ip)} -\frac{2W(t)}{\delta ip} +1 \Bigr)e^{iPt}\, .
\end{equation}
The corresponding modes $h_{n}(P)$ can be plugged into  (\ref{RG1}), (\ref{RGsecond}) to obtain 
series expansions for  $\mu(t,P,Q)$ and $\eta(t,P)$ in the variable $x=\nu e^{\delta t}$. One obtains  
\begin{eqnarray}\label{series_exps}
\mu(t,P,Q)&=&  \xi_{0}\delta(Q-|P|) + i\frac{f(Q)}{Q}\hat D_{p}(x)[ \Phi(-x,1,1+i(p-q)) -  \Phi(-x,1,1-i(p-q))] \, , 
\nonumber \\
\eta(t,P)&=& 2\hat D_{p}(x)[\Phi(-x,1,2 +ip) - \Phi(-x,1,ip)]
\end{eqnarray}
where $\hat D_{p}(x)$ is a differential operator 
\begin{equation}\label{Ddiff}
\hat D_{p}(x)= x + \frac{2x^{2}}{ip}\frac{d}{dx} + \frac{x^{3}}{ip(1+ip)}\frac{d^{2}}{dx^{2}} 
\end{equation}
and $\Phi(z,s,v)$ stands for the Lerch phi-function (see e.g. \cite{GR} section 9.55).
The last one is defined in the region $|z|<1$ and  for $v\ne 0,-1,-2,\dots$ by a    
power series expansion 
\begin{equation}
\Phi(z,s,v) = \sum_{n=0}^{\infty} \frac{z^{n}}{(n + v)^{s}} 
\end{equation}
 and is analytically 
extended to the complex plane with a branch cut going over the real line from $z=1$ to $z=+\infty$. 
For $s=1$, which is the case at hand, the  Lerch phi function can be expressed via the hypergeometric  ${}_2F_{1}$ function as 
$$
\Phi(z,1,v) = \frac{{}_2F_{1}(1,v,1+v;z)}{v}  \,.
$$
For future reference we record here the identity 
\begin{equation}\label{phi_asympt}
\Phi(z,1,v)  =  \frac{\pi}{\sin(\pi v)}(-z)^{-v}  + z^{-1}\Phi(z^{-1},1,1-v) 
\end{equation}
which can be used to obtain the asymptotic expansion near $z=\infty$.

We can thus conclude from (\ref{series_exps}), (\ref{Ddiff})  that although the perturbation series for the time-dependent couplings 
 $\mu(t,P,Q)$ and $\eta(t,P)$  is initially set up for sufficiently large and negative values of $t$, more precisely for 
 $t<-\delta^{-1}\ln \nu$,  it can be extended via analytic continuation in the variable $x$ to all values of $t$. In particular 
 one can study the $t\to +\infty$  asymptotic. 

Although the above representation of solutions via Lerch phi-function is important in establishing its convergence properties 
in practice we will find it more useful to work with their  spectral representations.
With the function $h(t,P)$ given explicitly in (\ref{hsol}) equations (\ref{diffeqs}) take the form of oscillator equations with a driving force. 
They are solved by passing to the Fourier transforms. Taking into account the initial conditions (\ref{ics}) we obtain  
\begin{eqnarray}\label{specs}
\mu(t,P,Q)&=& e^{ iPt}\xi_{0}\delta(|P|-Q) + f(Q)\xi_{0} \int\limits_{-\infty}^{+\infty} d\omega\, e^{-i\omega t} 
\frac{d_{P}(\omega)}{(\omega + i\epsilon)^{2} - Q^{2}} \, ,  \nonumber \\
\eta(t,P)&=&2\xi_{0}\delta \int\limits_{-\infty}^{+\infty} d\omega\, e^{-i\omega t} \frac{ d_{P}(\omega)}
{\delta^{2} + \omega^{2}}
\end{eqnarray}
where 
\begin{equation}\label{d}
d_{P}(\omega)=\frac{2\omega(1 + i\omega/\delta) }{\delta p(1+ip)}\hat W(\omega +P) 
\end{equation}
and
\begin{equation} \label{FW}
\hat W(\omega)=\frac{i\nu^{-i\omega/\delta}}{2\sinh[\pi(\omega  + i\epsilon)/\delta]}
\end{equation}
is the Fourier transform of $W(t)$.

Note that the corresponding operators $\Phi_{P}(\nu)$ defined in (\ref{newops}) satisfy the 
following hermitean conjugation rule
\begin{equation}
\Phi_{P}(\nu)^{\dagger}=\Phi_{-P}(\nu) \, . 
\end{equation}


\subsection{Zero momentum solutions}
The case $P=0$ needs special care. One notices that solutions (\ref{specs}) do not have 
a $P\to 0$ limit unless the normalization factor $\xi_{0}=\xi_{0}(P)$ vanishes at least as fast as $P$. 
We will see in the next section that a natural normalization factor for our solution is 
such that it vanishes only as $|P|^{1/2}$ that does not compensate  
 the blow up of the spectral function $d_{P}(\omega)$ in the $P\to 0$ limit. The physical reason for this 
apparent singularity is that at $P=0$ the RG equations for the coupling constants 
$u(t)$, $\lambda(q,t)$ are no longer independent from the $\eta(t)$, $\mu(t,Q)$ variables.

One can find two distinct solutions at zero momentum by taking  limits of suitable  linear combinations 
of solutions (\ref{specs}). We first consider a solution $\delta(\nu)$ defined as
\begin{equation}
\delta(\nu) \equiv \lim\limits_{P\to 0}P \Phi_{P}(\nu)
\end{equation} 
where we took the normalization factor $\xi_{0}$ to be identically one.
The corresponding coupling constants are
\begin{eqnarray}\label{delta_expl}
\mu_{\delta \nu}(t,Q) &=& f(Q)\int\limits_{-\infty}^{+\infty}\!\! d\tilde \omega\,  e^{-i\tilde \omega \delta t}
\frac{d_{\delta\nu}(\tilde \omega)}
{[(\tilde \omega + i\epsilon)^{2}-q^{2}]}\, ,   \nonumber \\
\eta_{\delta \nu}(t) &=& 2\delta \int\limits_{-\infty}^{+\infty}\!\! d\tilde \omega \,  e^{-i\tilde \omega \delta t}
\frac{d_{\delta\nu}(\tilde \omega)}
{(1 + \tilde \omega^{2})}\,  
\end{eqnarray}
 where $\tilde \omega = \omega/\delta$ and 
 \begin{equation}
 d_{\delta\nu}(\tilde \omega) = 2\tilde \omega(1 + i\tilde \omega)\hat W(\delta \tilde \omega)\, . 
 \end{equation}
 It is straightforward to check that this solution describes the marginal operator corresponding 
 to   variation of  the bare coupling constant $\nu$. One has 
 \begin{equation}
 \mu_{\delta \nu}(t,Q) = 2i\nu \frac{\partial}{\partial \nu} \lambda(Q,t) \, , \quad 
 \eta_{\delta \nu}(t) = 2i\nu \frac{\partial}{\partial \nu} U(t)
 \end{equation}
 where $\lambda(Q,t)$ and $U(t)$ were found in \cite{2}.
 The solution $\delta(\nu)$ asymptotes to zero at $t\to -\infty$:
 \begin{equation}
 \lim\limits_{t\to -\infty} \mu_{\delta \nu}(t,Q) = 0\, , \quad 
 \lim\limits_{t\to -\infty} \eta_{\delta \nu}(t) = 0\, . 
 \end{equation}
 The existence of such a solution means that one cannot fix a unique solution to equations 
 (\ref{diffeqs}) by fixing its leading asymptotic  at $t\to -\infty$. In order to fix a solution 
 uniquely  one should also fix the 
 first subleading terms in the series expansion in $e^{\delta t}$ for $t\ll 0$. It suffices to fix 
 the coefficient $\eta_{1}$ for that purpose.

Another zero momentum  solution which we denote $\Phi_{0}(\nu)$ 
corresponds to a perturbation at $t\to -\infty$ by a cosmological constant 
operator $\Phi_{0}^{\delta}$.  It can be obtained as a limit 
\begin{equation}\label{Phi_0}
\Phi_{0}(\nu)\equiv \lim\limits_{P\to 0}\frac{1}{2}(\Phi_{P}(\nu) + \Phi_{-P}(\nu))\, . 
\end{equation}
where the normalization factor $\xi_{0}$ is again chosen to be identically one. 
The coupling constants corresponding to (\ref{Phi_0}) read 
\begin{eqnarray}\label{Phi0_expl}
\mu_{0}(t,Q) &=& \delta(Q)  + \frac{f(Q)}{\delta}\int\limits_{-\infty}^{+\infty}\!\! d\tilde \omega\,  
e^{-i\tilde \omega \delta t} \frac{d_{0}(\tilde \omega)}{ [(\tilde \omega + i\epsilon)^{2}-q^{2}]}\, ,  \\
\eta_{0}(t) &=& 2\int\limits_{-\infty}^{+\infty}\!\! d\tilde \omega \,  e^{-i\tilde \omega \delta t} 
 \frac{d_{0}(\tilde \omega)}{(1 + \tilde \omega^{2})}
\end{eqnarray}
where 
\begin{eqnarray}\label{d0}
d_{0}(\tilde \omega) = &&\lim\limits_{P\to 0}\frac{1}{2}(d_{P}(\tilde \omega) + d_{-P}(\tilde \omega)) = \nonumber \\
&& -i(1 + \ln(\nu))d_{\delta \nu}(\tilde \omega) + 
\tilde \omega (\tilde \omega - i) 
\pi\frac{\nu^{-i\tilde \omega}\cosh(\pi\tilde \omega)}{\sinh^{2}[\pi(\tilde \omega + i\epsilon)]}\, . 
\end{eqnarray}

This solution can be simplified by subtracting a suitable amount of the previously found  solution $\delta(\nu)$ 
 so that the spectral function $d_{0}(\tilde \omega)$ is replaced by 
\begin{equation}
d_{0}'(\tilde \omega) \equiv
\tilde \omega (\tilde \omega - i) 
\pi\frac{\nu^{-i\tilde \omega}\cosh(\pi\tilde \omega)}{\sinh^{2}[\pi(\tilde \omega + i\epsilon)]}\, . 
\end{equation}
We denote the corresponding solution by $\Phi_{0}'(\nu)$. 


\section{The \inn and \out states}\label{in_out}
 \setcounter{equation}{0}
\subsection{Conformal primaries at the infrared fixed point}
The $t\to -\infty$ limit of  the time-dependent  BCFT  (\ref{interaction})  is the  unperturbed FZZT theory characterized by $\delta$. 
The incoming scattering states are  the states corresponding to the boundary operators $e^{iPX_{0}}\Phi_{|P|}^{\delta}$. As 
in the previous section we consider momenta $P= {\cal O}(\delta)$. 

It was shown in \cite{2} that the $t\to \infty$ limit  can be described by 
the $\delta_{*}=-\delta$ FZZT boundary condition perturbed by a marginal  oscillating term  (\ref{lambda_as}) interpreted as radiation. 
It is natural then to define the out-states as operators of the form  $e^{iPX_{0}}\Phi_{|P|}^{-\delta}$. 
Since the time-dependent  theory  (\ref{interaction})  is constructed as a perturbation in  the Fock space  (\ref{spectrum}) of the theory labeled by $\delta$ 
one needs to construct in that Fock space the operators  $\Phi_{|P|}^{-\delta}$ out of the operators $\Phi_{|P|}^{\delta}$, $\Phi_{\vartheta}^{\delta}$.
This problem is most elegantly solved by using the boundary RG flow generated by the relevant operator $ \Phi_{\vartheta}^{\delta}$.  
This RG flow  was studied in \cite{1}, \cite{2}.  The renormalized boundary action containing all near-marginal couplings generated by the flow has 
the form 
\begin{equation}
S_{\rm Bd}^{\rm ren} = \int_{\mathbb R}\! dx\, \Bigl(  u \epsilon^{-\delta^{2}} \Phi_{\vartheta}^{\delta}(x) + 
\int\limits_{0}^{\infty}\! dP\, \lambda(P) \epsilon^{P^{2}} \Phi_{P}^{\delta}(x) \Bigr)
\end{equation} 
where $\epsilon$ is the regularization scale. The RG equations computed in the $\delta \ll 1$ conformal perturbation theory 
read \cite{2}
\begin{eqnarray}\label{RGeqs}
\epsilon\frac{d u}{d\epsilon}&=&\delta^{2} u -2\delta w^{2}\, , \nonumber \\
 \epsilon\frac{d \lambda(P)} {d\epsilon}&=& -P^{2}\lambda(P) - \frac{2P^{2}}{\pi(\delta^{2} + P^{2})}w^{2}
\end{eqnarray}
where
$$
w= u + \int\limits_{0}^{\infty}\! dP\, \lambda(P) \, . 
$$
It was shown in \cite{2} that the infrared fixed point of the above equation 
$$
u_{*}=2\delta \, , \qquad \lambda_{*}(P) =  - \frac{2\delta^{2}}{\pi(\delta^{2} + P^{2})}
$$
corresponds to the $\delta_{*}=-\delta$ FZZT conformal boundary condition. 

In general if beta functions have the form:
\begin{equation}
\beta^j = \delta^{j}\lambda^{j} - \sum_{kl}C_{kl}^{j}\lambda^{k}\lambda^{l}
\end{equation}
and $\lambda^j=\lambda_*^j$ is an IR fixed point, then 
\begin{equation}
D_{i}^{j}\equiv (\partial_{i}\beta^{j})(\lambda_{*}) 
\end{equation}
is the matrix of anomalous dimensions at the IR fixed point. Its spectrum of eigenvalues 
gives the IR fixed point anomalous dimensions and its eigenvectors the expressions for IR 
fixed point conformal primaries via the UV ones\footnote{Of course this can be used to obtain 
only a subset  of the  primaries  which are tangential to the RG flow in the vicinity of the IR fixed point} 
(which is sensible in the small $\delta$ expansion).

For the model at hand we find using (\ref{RGeqs}) that the operator playing the role of the matrix $D_{i}^{j}$ 
above acts  as  
\begin{eqnarray} \label{D}
\hat D\Phi_{\vartheta}^{\delta} = \delta^{2}\Phi_{\vartheta}^{
\delta}  -2\delta
(\int\limits_{0}^{\infty}\! dQ\, f(Q) \Phi_{Q}^{\delta}
+2\delta \Phi_{\vartheta}^{\delta})&& 
\nonumber \\
\hat D\Phi_{P}^{\delta} = -P^{2}\Phi_{P}^{\delta}  -2\delta 
(\int\limits_{0}^{\infty}\! dQ\, f(Q) \Phi_{Q}^{\delta}  +2\delta \Phi_{\vartheta}^{\delta} )&&
\end{eqnarray}
where $f(Q)$ is given in (\ref{f}).

To find the eigenvectors of  operator $\hat D$ we write an ansatz
\begin{equation}\label{Phst}
\Phi_{P}^{*} = \Phi_{\vartheta}^{\delta} + \int\limits_{0}^{\infty}\! dQ\, K_{P}(Q)\Phi_{Q}^{\delta} \, . 
\end{equation}
The eigenvector equation
\begin{equation} \label{DPh}
\hat D \, \Phi_{P}^{*} = - P^{2}\, \Phi_{P}^{*} 
\end{equation} 
 implies
\begin{eqnarray} \label{eqs}
K_{P}(Q)(Q^{2}-P^{2}) = -\frac{\delta}{2}(p^{2} + 1) f(Q) \, , && \nonumber \\
1 + \int\limits_{0}^{\infty}\! dQ \, K_{P}(Q) = \frac{1}{4}(p^{2}+1) \, .&&
\end{eqnarray}

We find a distributional solution  
\begin{equation}\label{K+}
K_{P}(Q)= g_{p}\delta(P-Q)  - \frac{\delta(p^2 +1)f(P)}{2(Q^{2}-P^{2} + i\epsilon P )}
\end{equation}

The coefficient $g_{p}$ is determined by substituting (\ref{K+}) into the second equation in 
(\ref{eqs}). We obtain 
\begin{equation} \label{g}
g_{p}  = \frac{1}{4}(p-i)^{2} \, . 
\end{equation}

We also checked that there are no solutions to the eigenvalue equation (\ref{DPh}) for imaginary $P$ 
which is consistent with the fact that in the IR fixed point there are no normalizable relevant operators. 

We can check  that the wavefunctions of operators (\ref{K+}), (\ref{g}) have the correct asymptotics at  
$\phi_{0} \to -\infty$. 
Assuming the $\phi_{0}\to -\infty$ limit can be taken inside the integral over $P$ in  (\ref{Phst}) 
 using (\ref{Psi_asympt}) and taking the appropriate residues we obtain the following asymptotic
\begin{eqnarray}
\Psi_{\vartheta}(\phi_{0}) + \int\limits_{0}^{\infty}\! dQ\,  K_{p}(Q) \Psi_{Q}(\phi_{0}) 
\sim \frac{1}{4}[(p+i)^{2}C_{\delta}(P)e^{iP\phi_{0}} 
+ (p-i)^{2}C_{\delta}(-P)e^{-iP\phi_{0}}]= &&\nonumber \\
(p^{2}+1)\frac{1}{4}[C_{-\delta}(P)e^{iP\phi_{0}} 
+ C_{-\delta}(-P)e^{-iP\phi_{0}}]&&
\end{eqnarray}

This means that to the leading order in $\delta$ we can identify
\begin{equation}\label{phi_new}
\Phi^{-\delta}_{P} = \frac{4}{p^{2}+1}\Phi_{P}^{*}= \frac{4}{p^2 +1}\Phi_{\vartheta}^{\delta} + \left(\frac{p-i}{p+i}\right)\Phi_{|P|}^{\delta} 
-  2\delta \int\limits_{0}^{\infty}\! dQ \frac{f(Q)}{(Q^{2}-P^{2} + i\epsilon P)}\Phi_{Q}^{\delta} \, . 
\end{equation}
Note the following simple  properties  
$$
\Phi^{\pm \delta}_{P}= \Phi^{\pm \delta}_{-P} = (\Phi^{\pm \delta}_{P})^{\dagger}\, . 
$$


\subsection{Conserved inner product and normalizations} 

An important property of the time evolution  equations (\ref{diffeqs}) is the existence of a conserved Klein-Gordon type 
inner product. For any two  solutions $(\mu_{i}, \eta_{i}), i=1,2$ we define the inner product to be 
\begin{eqnarray}\label{iproduct}
&&\langle (\mu_{1},\eta_{1}), (\mu_{2},\eta_{2})\rangle_{KG}=\frac{i}{2}\int\limits_{0}^{\infty}\!\! dQ\,  C_{\delta}(Q)C_{\delta}(-Q) 
(\mu_{2}^{*}\partial_{t}\mu_{1} - 
  \mu_{1}\partial_{t}\mu_{2}^{*})  + \frac{i}{2}d_{\delta}(\eta_{2}^{*}\partial_{t}\eta_{1} - \eta_{1}\partial_{t}\eta_{2}^{*})  =\nonumber \\
 &&\frac{i\pi \mu_{r}}{2\delta }\Bigl[ \int\limits_{0}^{\infty}\!\! dQ \frac{\pi \delta (\delta^{2} + Q^{2})}{Q^{2}}
(\mu_{2}^{*}\partial_{t}\mu_{1} - 
  \mu_{1}\partial_{t}\mu_{2}^{*})  + (\eta_{2}^{*}\partial_{t}\eta_{1} - \eta_{1}\partial_{t}\eta_{2}^{*}) \Bigr] \, .
  \end{eqnarray}
It is straightforward to check using (\ref{diffeqs}) that this inner product is $t$-independent. The weight function under the integral 
and the coefficient at the term with $\eta_{1,2}$  can be understood as a consequence of the two-point function normalizations (\ref{2p1}), (\ref{2p2}). 
We will use the same notation $\langle \Phi_{P_{1}}(\nu) , \Phi_{P_{2}}(\nu) \rangle_{KG}$ to define the inner product for operators (\ref{newops}) 
corresponding  to solutions of (\ref{diffeqs}), (\ref{ics}). 

The operators $e^{iPX_{0}}\Phi_{|P|}^{\delta}$ are asymptotic solutions to (\ref{diffeqs}) in the far past. 
Since the inner product (\ref{iproduct}) is conserved  it can be evaluated on such operators. 
We   choose a basis of normalized positive frequency  \inn states to be given  by operators 
\begin{equation}\label{in_sts}
{\cal O}^{\rm in}_{P} = \frac{1}{|P|^{1/2}C_{\delta}(P)}\Phi^{\delta}_{P}e^{-iPX_{0}} \, , \enspace P>0 \, . 
\end{equation}
The hermitean conjugated operators  form the basis of negative frequency \inn states: 
\begin{equation}
{\cal O}^{\rm in *}_{P} = \frac{1}{|P|^{1/2}C_{\delta}(-P)}\Phi^{\delta}_{P}e^{iPX_{0}} \, , \enspace P>0 \, .
\end{equation}
These operators satisfy the orthogonality conditions 
\begin{equation} \label{inin}
\langle {\cal O}^{\rm in}_{P_1}, {\cal O}^{\rm in}_{P_2}\rangle_{KG} = 
\delta(P_{1}- P_{2})\, , \quad 
\langle {\cal O}^{\rm in *}_{P_1}, {\cal O}^{\rm in *}_{P_2}\rangle_{KG} = 
-\delta(P_{1}- P_{2})\, , \quad 
\langle {\cal O}^{\rm in *}_{P_1}, {\cal O}^{\rm in}_{P_2}\rangle_{KG} = 0 \, . 
\end{equation}

We are next going to show that operators $\Phi^{-\delta}_{|P|}e^{iPX_{0}}$ are asymptotic solutions 
to (\ref{diffeqs}) at $t\to +\infty$. Substituting the asymptotic value $\delta =\lim_{t\to +\infty} W(t)$ for $W(t)$ 
 we rewrite   (\ref{diffeqs})  in the following form 
 \begin{eqnarray}
 \partial^{2}_{t} \mu(t,P,Q) &=& \hat D \mu (t,P,Q) \equiv  -Q^{2}\mu(t,P,Q) -2\delta f(Q) (\eta(t,P) + \int\limits_{0}^{\infty}\! dQ' \mu(t,P,Q')) \, , 
 \nonumber \\
 \partial^{2}_{t} \eta(t,P) &=& \hat D \eta(t,P)   \equiv  \delta^{2}\eta(t,P) -4\delta f(Q) (\eta(t,P) + \int\limits_{0}^{\infty}\! dQ' \mu(t,P,Q')) \, .
  \end{eqnarray}
 The operator $\hat D$ defined on the right hand sides can be recognized as the dual action on the coupling constants of 
 the dilatation operator (\ref{D}). Thus by (\ref{DPh}), (\ref{phi_new}) the operators $\Phi^{-\delta}_{|P|}e^{iPX_{0}}$ are indeed  solutions 
of the wave equation  (\ref{diffeqs}) at $t\to +\infty$. 
 
 We define a basis of positive frequency \out states 
 \begin{equation}\label{out_sts}
 {\cal O}^{\rm out}_{P}= \left(\frac{p-i}{p+i}\right)\frac{1}{|P|^{1/2}C_{\delta}(P)}\Phi^{-\delta}_{P}e^{-iPX_{0}} 
 \, , \enspace P>0 
 \end{equation}
 and a hermitean conjugated basis ${\cal O}^{\rm out *}_{P}$ of negative frequency states.
 The normalization factor is chosen so that 
 \begin{eqnarray}\label{complete_out}
& \langle {\cal O}^{\rm out}_{P_1}, {\cal O}^{\rm out}_{P_2}\rangle_{KG} = 
\delta(P_{1}- P_{2})\, , \quad 
\langle {\cal O}^{\rm out *}_{P_1}, {\cal O}^{\rm out *}_{P_2}\rangle_{KG} = 
-\delta(P_{1}- P_{2})\, , &\nonumber \\
& \langle {\cal O}^{\rm out *}_{P_1}, {\cal O}^{\rm out}_{P_2}\rangle_{KG} = 0 \, . &
\end{eqnarray} 
This is  checked by  direct computation substituting the coefficients from  (\ref{phi_new}) into  (\ref{iproduct}).
The particular choice of phase in (\ref{out_sts}) is done for convenience.

\subsection{The $t\to +\infty$ asymptotics}

One can compute the $t\to +\infty$ asymptotics of the  time-dependent vertex operators 
(\ref{newops}). What we formally mean by taking such a limit is taking first the limit of the 
analytically continued solutions to the wave equations (\ref{series_exps}). The asymptotic coupling 
constants proportional to $ e^{iQt}$ can then be coupled to operators $e^{iQX_{0}}\Phi_{Q'}^{\delta}$ giving 
thus the operator-valued  limit\footnote{ Putting it a bit differently (c.f. the discussion in  \cite{2}) 
 one can speak of a limit for the wave functions obtained as an overlap of the states corresponding to (\ref{newops}) 
with the states $\langle \phi_{0}^{op} |\otimes \langle t | $ as in (\ref{wfns}) and then invoke the state-operator correspondence.}. 
The desired asymptotic can be obtained either by using the appropriate  asymptotic expansion 
for the Lerch phi-function (hypergeometric function)  or by taking 
residues in the complex $\omega$-plane in the spectral formulas (\ref{specs}).    
One obtains 
\begin{eqnarray}\label{asymp1}
&& \mu(t,P,Q)  \underset{t \to  \infty }{\sim}  e^{ iPt}\xi_{0}\Bigl[\delta(P-Q) - \frac{2f(Q)\delta}{
(Q^{2} - P^{2} + i\epsilon P)}\left(\frac{p+i}{p-i}  \right) \Bigr] + 
\nonumber \\
&& \frac{\xi_{0}\pi f(Q)}{\delta p(1+ip)}\Bigl[   
e^{-iQt}\frac{ (1 + iq)\nu^{-i(q +  p)}}{ \sinh[\pi(q +  p -i\epsilon)]} 
   + e^{iQt}\frac{ (1 - iq) \nu^{-i(  p-q)}}{ \sinh[\pi(-q +  p - i\epsilon)]} \Bigr]\, .
\end{eqnarray}
Note that when working with the first spectral expression (\ref{specs}) we have to take into account 
two $i\epsilon$ contour prescriptions: one explicitly present in (\ref{specs}) and another one 
in the Fourier transform of $W$ (\ref{FW}). The $t\to \infty$ limit  depends on the difference of two 
epsilons. The choice of the sign is however insignificant as the complete expression is   
 non-singular at $p=q$. The particular choice of sign was made for an easy comparison with (\ref{phi_new}). 
Similarly we obtain 
\begin{equation}\label{asymp2}
\eta(t,P)  \underset{t \to  \infty }{\sim}  \frac{4}{(p-i)^{2}}e^{ iPt}\, .
\end{equation}

Comparing (\ref{ics}) with (\ref{in_sts}) we can fix the normalization constant 
\begin{equation}\label{xi0}
\xi_{0}=\xi_{0}(P)=\frac{1}{|P|^{1/2}C_{\delta}(-P)}
\end{equation}
so that in the far past the operator $\Phi_{-P}(\nu)$, $P>0$ approaches the normalized 
positive frequency in operator ${\cal O}^{\rm in}_{P}$: 
\begin{equation}\label{Phi_in}
\Phi_{-P}(\nu) \underset{t \to  -\infty }{\sim} {\cal O}^{\rm in}_{P}\, , \enspace P>0\, . 
\end{equation}

From now on we will assume that $P>0$ and will work with the vertex operator $\Phi_{-P}(\nu)$. 
Combining (\ref{asymp1}) and (\ref{asymp2}) together and taking into account 
 (\ref{phi_new}), (\ref{out_sts}) we obtain 
 \begin{eqnarray}\label{Phi_asympt}
 &&\Phi_{-P}(\nu)  \underset{t \to  \infty }{\sim}  {\cal O}^{\rm out}_{P}  + 
 \frac{\xi_{0}(-P)\pi }{\delta p(1-ip)}\times \nonumber \\
 &&   \int\limits_{0}^{\infty}\!\! dQ  f(Q) \Phi_{Q}^{\delta} 
 \Bigl[e^{-iQX_{0}}\frac{ (1 + iq)\nu^{i(p-q)}}{ \sinh[\pi(p-q + i\epsilon)]} 
   + e^{iQX_{0}}\frac{ (1 - iq) \nu^{i(  p+q)}}{ \sinh[\pi(  p + q + i\epsilon)]}\Bigr]\, .
   \end{eqnarray}

 For the zero momentum solutions $\delta(\nu)$ and $\Phi_{0}(\nu)$ 
 one finds the following asymptotics 
 \begin{eqnarray} \label{zero_asympt}
 && \delta(\nu) \underset{t \to  \infty }{\sim}
  \int\limits_{0}^{\infty}\!\! dQ  \frac{2q^{2}\Phi_{Q}^{\delta} }{\sinh(\pi q)} 
 \Bigl[\frac{ e^{-iQX_{0}}\nu^{-iq}}{ 1-iq} -
    \frac{ e^{iQX_{0}}\nu^{i q}}{ 1 +iq}\Bigr]\, , \\
 &&\Phi_{0}(\nu) \underset{t \to  \infty }{\sim} -\Phi_{0}^{-\delta} 
 - \nonumber \\
 &&\frac{2\pi }{\delta}\int\limits_{0}^{\infty}\!\! dQ  q^{2}\cosh(\pi q)\Phi_{Q}^{\delta}  
 \Bigl[\frac{ e^{-iQX_{0}}\nu^{-iq}}{ (1-iq)\sinh^{2}[\pi(q+i\epsilon)]} +
    \frac{ e^{iQX_{0}}\nu^{i q}}{ (1 +iq)\sinh^{2}[\pi(q-i\epsilon)]}\Bigr] \, . 
 \end{eqnarray}
 
 
\section{Solutions asymptoting to  positive frequency out states}\label{solutions2}
\setcounter{equation}{0}
The $t \to \pm \infty$ asymptotics of the zero mode $\delta(\nu)$ present us the following 
apparent problem. On the one hand the $t\to + \infty$ asymptotic given in (\ref{zero_asympt}) means 
that the asymptotic overlaps $$\langle {\cal O}^{\rm out}_{Q}, \delta(\nu)\rangle_{KG}$$ are nonvanishing. 
On the other hand given a solution $\Psi_{-P}(\nu)$ that at $t\to + \infty$ approaches ${\cal O}^{\rm out}_{P}$, 
its overlap with $\delta(\nu)$ would be zero provided the $t\to -\infty$ asymptotic of $\Psi_{-P}(\nu)$
is bounded. This is because the $\delta(\nu)$ solution asymptotes to zero at $t\to -\infty$. 
We are forced to conclude that either the solutions approaching    ${\cal O}^{\rm out}_{P}$ in the infinite 
future do not exist at all or, if they exist, they blow up at $t\to -\infty$. We will show in this section 
that it is the second option that is realized.  
 
A solution $\Psi_{P}(\nu)$ that  asymptotically approaches  ${\cal O}^{\rm out *}_{P}$ 
can be constructed as follows. We start by looking at a solution at large positive values 
of $t$ in the form of series expansions
\begin{equation} \label{future_exp}
\tilde \mu(t,Q,P) = \sum_{n=0}^{\infty} \tilde \mu_{n} e^{(iP -n\delta)t} \, , 
\quad \tilde \eta(t,P) = \sum_{n=0}^{\infty} \tilde \eta_{n} e^{(iP -n\delta)t}
\end{equation} 
with 
\begin{equation}\label{mu0}
\tilde \mu_{0}=\frac{4\tilde h_{0}}{p^{2}+1}K_{P}(Q)=\tilde h_{0}\Bigl[
\left(\frac{p-i}{p+i}\right)\delta(Q-|P|) -\frac{2\delta f(Q)}{Q^{2}-P^{2} + i\epsilon P}
\Bigr]\, ,  
\end{equation}
\begin{equation}
\tilde \eta_{0}=\frac{4\tilde h_{0}}{p^{2}+1}
\end{equation}
where $\tilde h_{0}$ is a normalization factor and $P>0$. 
The coefficients $\tilde \mu_{0}$, 
$\tilde \eta_{0}$ above are chosen so that 
\begin{equation}
\Psi_{P}(\nu) \equiv \tilde \eta(t,P) \Phi_{\vartheta}^{\delta} 
+ \int\limits_{0}^{+\infty}\!\! dQ\, \tilde \mu(t,Q,P)\Phi_{Q}^{\delta} \underset{t \to  \infty }{\sim} 
\tilde h_{0}\Phi_{|P|}^{-\delta}e^{iPt} \, .
\end{equation}
Choosing the normalization factor 
\begin{equation}
\tilde h_{0}=\tilde h_{0}(P) = \left( \frac{p+i}{p-i} \right)\frac{1}{|P|^{1/2}C_{\delta}(-P)}
\end{equation}
we obtain
\begin{equation}
\Psi_{P}(\nu) \underset{t \to  \infty }{\sim} {\cal O}^{\rm out *}_{P}\, , \quad 
\Psi_{-P}(\nu) \underset{t \to  \infty }{\sim} {\cal O}^{\rm out }_{P} \, .
\end{equation}

Substituting the expansions (\ref{future_exp}) into the wave equations (\ref{diffeqs}) 
we obtain the following set of equations 
\begin{eqnarray} \label{fme}
&& [Q^{2} + (iP-n\delta)^{2}]\tilde \mu_{n} = -2f(Q)\sum\limits_{l=0}^{n}\tilde W_{l}
\tilde h_{n-l} \, ,  \nonumber \\
&& [ (iP-n\delta)^{2} - \delta^{2}] \tilde \eta_{n} = -4\delta 
\sum\limits_{l=0}^{n}\tilde W_{l}
\tilde h_{n-l}
\end{eqnarray}
where 
\begin{equation}
\tilde h_{n}=\tilde \xi_{n} + \tilde \eta_{n}\equiv \int\limits_{0}^{+\infty}\!\! dQ\, \mu_{n}(Q) + \tilde \eta_{n}
\end{equation}
and 
$\tilde W_{l}=\delta (-\nu)^{-l}$ are coefficients in the expansion 
$$
W(t) = \sum_{l=0}^{\infty} \tilde W_{l}e^{-n\delta t} \, . 
$$

We solve the system (\ref{fme}) by similar steps to those we did in section 3.2 solving  (\ref{RGadd}).
The first equation in (\ref{fme}) together with (\ref{mu0}) imply
\begin{equation}
\tilde \mu_{n} = -\frac{2f(Q)(W\cdot h)_{n}}{Q^{2} + (i(P-i\epsilon)-n\delta)^{2}} 
+ \delta_{n,0}\tilde h_{0}\left(\frac{p-i}{p+i}\right)\delta(Q-|P|)
\end{equation}
where for brevity we used the notation 
$$
(W\cdot h)_{n}\equiv \sum_{l=0}^{n}\tilde W_{l}
\tilde h_{n-l} \, . 
$$
Integrating the last equation over $Q$  we obtain 
\begin{equation}\label{txi}
(\delta(n+1) -iP)\tilde \xi_{n} = -2(W\cdot h)_{n} -i\delta_{n,0}\tilde h_{0}\frac{(p+i)^{2}}{p-i}\delta \, . 
\end{equation}
Taking the appropriate linear combinations of equations (\ref{fme}), (\ref{txi}) we obtain 
\begin{eqnarray} \label{modes2}
&&\tilde \mu_{n} = \frac{\delta f(Q)\tilde h_{n}(n-1+ip)}{Q^{2} + (iP +\epsilon -n\delta)^{2}} + 
\delta_{n,0}[\tilde \mu_{0} + \frac{\delta f(Q)\tilde h_{0}(1-ip)}{Q^{2}- (iP+\epsilon)^{2}}] 
\, , \nonumber \\
&& \tilde \eta_{n}= \frac{2\tilde h_{n}}{n+1 -ip} -\frac{2i}{p-i}\tilde h_{0}\delta_{n,0}\, , 
\end{eqnarray}
\begin{equation}
(iP+\delta(1-n))\tilde h_{n} = 2(W\cdot h)_{n} + i\delta_{n,0}\tilde h_{0}(p+i)\delta \, . 
\end{equation}
The last equation is solved in terms of hypergeometric series
\begin{equation}
\tilde h(t) = \tilde h_{0}\Bigl[ \left(\frac{p+i}{p-i} \right) -\frac{2i}{p-i}{}_{2}F_{1}(1,-1-ip,2-ip; -\tilde x) \Bigr]e^{iPt} \, . 
\end{equation}
where $\tilde x =\nu^{-1}e^{-\delta t}$.
 Substituting the  coefficients $\tilde h_{n}$ into (\ref{modes2}) 
 we obtain the following expressions for $\tilde \mu(t,Q,P)$, $\tilde \eta(t,P)$ in 
terms of Lerch phi-functions 
\begin{eqnarray}\label{tsol1}
&& \tilde \mu(t,Q,P) = \delta(Q-|P|)\tilde h_{0} \left(\frac{p-i}{p+i} \right)e^{iPt} + 
\tilde h_{0}\frac{f(Q)}{Q} ip(p+i)e^{iPt}\Bigl[ -\frac{\Phi(-\tilde x, 1,i(q-p))}{q(q+i)} - \nonumber \\
&& \frac{2i}{q}\Phi(-\tilde x,1,-ip) + \frac{\Phi(-\tilde x,1,-i(q+p))}{q(q-i)} + \frac{2iq}{1+q^{2}}\Phi(-\tilde x,1,1-ip)\Bigl]\, , 
\end{eqnarray}
\begin{eqnarray}\label{tsol2}
 &&\tilde \eta(t,P)=  4\tilde h_{0}p(p+i)e^{iPt} 
 \Bigl[ (\frac{1}{2}\Phi(-\tilde x,2,1-ip) - \Phi(-\tilde x,1,-ip) + \nonumber \\
 &&  \frac{3}{4}\Phi(-\tilde x,1,1-ip) + \frac{1}{4}\Phi(-\tilde x,1,-1-ip)\Bigr] \, . 
\end{eqnarray}

An  asymptotic expansion of $\Psi_{P}(\nu)$ in the $t\to -\infty$ region can be obtained 
from  expressions (\ref{tsol1}), (\ref{tsol2}) using relation (\ref{phi_asympt}) for 
the functions $\Phi(-\tilde x,1,v)$ and the following integral expression 
\begin{equation} \label{Phi2}
\Phi(-\tilde x,2,1-ip)=\frac{1}{2i}\int\limits_{-\infty}^{+\infty}\!\!d\omega 
\frac{\nu^{-i\omega}e^{-i\omega\delta t}}{(\omega - p - i)^{2}\sinh[\pi(\omega + i\epsilon)]}\, .  
\end{equation}

   The leading terms in the  asymptotic read
\begin{eqnarray} \label{tmu_as}
&&\tilde \mu(t,Q,P) \underset{t \to  -\infty }{\sim} \tilde h_{0}\left(\frac{p-i}{p+i}\right)\delta(Q-|P|)e^{iPt} - \nonumber \\
&&\tilde h_{0} \pi p(p+i)\frac{f(Q)}{\delta q^{2}}\Bigl[ \frac{ \nu^{i(q-p)}e^{iQt}}{(q+i)\sinh[\pi(q-p)]}  +  
 \frac{ \nu^{-i(q+p)}e^{-iQt}}{(q-i)\sinh[\pi(q+p)]} \Bigr] + \nonumber \\
 && -2\frac{f(Q)}{\delta q^{2}}\alpha(p) \, . 
\end{eqnarray}
\begin{equation} \label{teta_as}
\tilde \eta(t,P) \underset{t \to  -\infty }{\sim} \alpha(p)
 (\nu^{-1}e^{-\delta t} + 4 + {\cal O}(e^{\delta t}) ) \, 
\end{equation}
where 
\begin{equation} \label{alphap}
\alpha(p) = \tilde h_{0}(P)\frac{\pi p(1-ip)\nu^{-ip}}{\sinh(\pi p)} 
=-i\left(\frac{p+i}{p-i} \right)\frac{|p|^{3/2} \nu^{-ip}}{(\mu_{r}\delta)^{1/2}\sinh(\pi p)}\, . 
 \end{equation}
The first term in (\ref{teta_as}) is exponentially divergent in the $t\to -\infty$ limit 
and therefore has a nonzero overlap with the $\delta(\nu)$ solution. This overlap matches with the 
one that can be computed from the expansions in the    $t\to +\infty$ region  using (\ref{teta_as}).


\section{Bogolyubov transformations}\label{Bogolyubov}
\setcounter{equation}{0}

To obtain the Bogolyubov coefficients we start by  recasting  asymptotics \ref{Phi_asympt} into a form 
\begin{equation}\label{Phi_out}
\Phi_{-P}(\nu)  \underset{t \to  \infty }{\sim}  \int\limits_{0}^{\infty}\! dQ 
( \beta_{P}(Q){\cal O}^{\rm out *}_{Q} + 
 \alpha_{P}(Q) {\cal O}^{\rm out}_{Q} ) 
 \end{equation}
where $\beta_{P}(Q)$ and $\alpha_{P}(Q)$ are Bogolyubov coefficients.
Using  the completeness of the out-basis (\ref{out_sts}) 
and the inner products (\ref{complete_out}) we can evaluate the Bogolyubov coefficients at hand 
via  the asymptotics of the inner-products 
\begin{equation}\label{alpbeta}
\alpha_{P}(Q) \underset{t \to  \infty }{\sim} \langle \Phi_{-P}(\nu), {\cal O}^{\rm out}_{Q} \rangle_{KG} \, , \quad 
\beta_{P}(Q) \underset{t \to  \infty }{\sim} -\langle \Phi_{-P}(\nu), {\cal O}^{\rm out *}_{Q} \rangle_{KG} \, . 
\end{equation}
It is technically simpler to compute overlaps of ${\cal O}^{\rm out}_{Q}$ with the asymptotic 
(\ref{Phi_asympt})
dropping any additional terms exponentially suppressed as $t\to \infty$. 
After the  substitution of (\ref{Phi_asympt}), (\ref{out_sts}) into (\ref{iproduct}) the computation 
reduces to finding the asymptotic value of the following integral 
\begin{equation}
I\equiv \int\limits_{-\infty}^{+\infty}\! ds \frac{s^{2}(q-s)}{(1-is)(s^{2}-q^{2} \pm i \epsilon)} 
\frac{e^{-i(s+q)\delta t}\nu^{i(p-s)}}{\sinh[\pi(s-p-i\epsilon')]} 
\end{equation}
where $+i\epsilon$ occurs in the computation of $\alpha_{P}(Q)$ and $-i\epsilon$ occurs for 
$\beta_{P}(Q)$. 
For $t\to +\infty$ this integral can be evaluated by taking residues in the region ${\rm Im} s \le 0$. 
Contributions from the residues located away from the real line are suppressed  
as $\exp(-n\delta t)$ and should be dropped. 
Evaluating the residues on the real line eventually yields 
the following expressions for the Bogolyubov coefficients 
\begin{eqnarray} \label{B_coefs}
\alpha_{P}(Q)&=& \delta(P-Q) - \frac{2|q|^{3/2}}{\delta |p|^{1/2}(p^{2}+ 1)}
\left( \frac{q+i}{q-i}\right) \frac{\nu^{i(p-q)}}{\sinh[\pi(p-q +i\epsilon)]} \, , \nonumber \\
\beta_{P}(Q) &=& \frac{2|q|^{3/2}}{\delta |p|^{1/2}(p^{2}+ 1)}
\left( \frac{q-i}{q+ i}\right) \frac{\nu^{i(p+q)}}{\sinh[\pi(p + q +i\epsilon)]}\, . 
\end{eqnarray}

Conservation of the inner product (\ref{iproduct}) together with asymptotics 
(\ref{Phi_in}), (\ref{Phi_out}) and formula (\ref{inin}) imply the following 
 pair of  relations for Bogolyubov coefficients 
\begin{eqnarray}\label{B_rels}
&& \int\limits_{0}^{\infty}\! dQ [ \alpha_{P_{1}}(Q)\alpha^{*}_{P_{2}}(Q) - 
\beta_{P_{1}}(Q)\beta^{*}_{P_{2}}(Q)] = \delta(P_{1}-P_{2}) \, , \nonumber \\
&& \int\limits_{0}^{\infty}\! dQ [ \alpha_{P_{1}}(Q)\beta_{P_{2}}(Q) - 
\alpha_{P_{2}}(Q)\beta_{P_{1}}(Q)] = 0 \, . 
\end{eqnarray}
These relations can be checked for the coefficients (\ref{B_coefs}) by 
direct computation. The details of this computation are relegated to  Appendix A.

A direct computation leads to another pair of relations
\begin{eqnarray}\label{B_rels2}
&& \int\limits_{0}^{\infty}\! dP [ \alpha_{P}^{*}(Q_1)\alpha_{P}(Q_2) - 
\beta_{P}(Q_1)\beta^{*}_{P}(Q_2)] = \delta(Q_{1}-Q_{2}) + d(Q_{1}, Q_{2}) \, , \nonumber \\
&& \int\limits_{0}^{\infty}\! dP [ \alpha_{P}^{*}(Q_{1})\beta_{P}(Q_{2}) - 
\alpha^{*}_{P}(Q_2)\beta_{P}(Q_1)] = -d(Q_{1}, -Q_{2})  
\end{eqnarray}
where an explicit expression for $d(Q_{1},Q_{2})$ is given in (\ref{dqq}). 
This pair of relations deviates from the ones one could have expected naively - the ones 
in which the function $d(Q_{1},Q_{2})$ 
is identically zero. This signals incompleteness of the set of incoming solutions $\Phi_{-P}(\nu)$.
 Let us recall the logic of derivation of the standard relations with vanishing  
$d(Q_{1},Q_{2})$. The inner products (\ref{alpbeta}) and the completeness of the asymptotic 
solutions ${\cal O}^{\rm out}_{P}$ imply that the solution 
\begin{equation}\label{chi}
\chi_{Q}(\nu)\equiv \int\limits_{0}^{\infty}\!\!dP\, [\alpha^{*}_{P}(Q)\Phi_{-P}(\nu) - \beta_{P}(Q)\Phi_{P}(\nu)] 
\end{equation} 
asymptotically at $t\to +\infty$ has the same overlaps with all $\Phi_{-P}(\nu)$ as ${\cal O}^{\rm out}_{Q}$. 
If the set of solutions $\Phi_{-P}(Q)$ was complete we would conclude that the solution 
$\chi_{Q}(\nu)$ asymptotically approaches ${\cal O}^{\rm out}_{Q}$ at $t\to +\infty$ (and thus 
 should  be identified with the solution $\Psi_{-P}(\nu)$ considered before). The inner product conservation 
would then imply $\langle \chi_{Q_1}(\nu), \chi_{Q_2}(\nu)\rangle_{KG}=\delta(Q_{1}-Q_{2})$, 
$ \langle \chi_{Q_1}^{*}(\nu), \chi_{Q_2}(\nu)\rangle_{KG}= 0$ which in its turn
would imply the standard relations (\ref{B_rels2}) with vanishing $d(Q_{1},Q_{2})$. 
We do know however from the considerations in section \ref{solutions2} that the set of solutions $\Phi_{-P}(Q)$ 
is incomplete due to the existence of solutions blowing up in the infinite past. 
In that section we constructed the solution $\Psi_{-P}(\nu)$ that approaches ${\cal O}^{\rm out}_{P}$
explicitly and saw that its asymptotic in the infinite past blows up. 

The complete asymptotic expansion in the $t\to -\infty$ region obtained using (\ref{phi_asympt}), 
(\ref{Phi2}) together with the asymptotic expansion for $\chi_{P}(\nu)$ in the same region can be used  
to identify the $\Psi_{-P}(\nu)$ solution with the following  linear combination 
of solutions 
\begin{equation}\label{outfromin}
\Psi_{-P}(\nu) = \chi_{P}(\nu) + \alpha(-p)\tilde \delta(\nu) + \beta(-p)\delta(\nu)
\end{equation} 
 where 
 \begin{equation}\label{tildedelta}
 \tilde \delta(\nu) = \Phi'_{0}(\nu) + \Phi_{\vartheta}^{\delta}\nu^{-1}e^{-\delta X_{0}}  
 +\Phi_{0}^{-\delta} \, , 
 \end{equation}
 \begin{equation}\label{betap}
 \beta(p)= -\frac{\pi}{\delta}\alpha(p)\coth(\pi p)\, . 
 \end{equation}
The above expansion means that in order to obtain a complete set of  scattering \out states 
one needs to include, in addition to solutions constructed in section \ref{vops}, a single
solution $\tilde \delta(\nu)$ blowing up in the $t\to -\infty$ limit.
 
We define  two  sets of solutions 
$$
{\cal S}^{\rm in}=\{ \Phi_{P}(\nu)\, , \enspace P\ne 0; \delta(\nu); 
\tilde \delta(\nu)\}\, , \quad  
  {\cal S}^{\rm out}=\{\Psi_{P}(\nu)\, , \enspace P\in {\mathbb R} \} \, . 
  $$ 
The inner products between 
the  bases of solutions asymptoting to the \inn and \out states are 
\begin{eqnarray} \label{innerprs}
&& \langle \Phi_{P}(\nu), \Phi_{Q}(\nu)\rangle_{\rm KG}=-{\rm sgn} (P) \delta(P-Q)\, , \nonumber \\
&& \langle \Psi_{P}(\nu), \Psi_{Q}(\nu)\rangle_{\rm KG}=-{\rm sgn} (P)\delta(P-Q) \, , \nonumber \\
&& \langle \Phi_{P}(\nu), \delta(\nu)\rangle_{\rm KG} = 
\langle \Phi_{P}(\nu), \tilde \delta(\nu)\rangle_{\rm KG} = 0\,  \enspace \, \mbox{for} \enspace P\ne 0 
\nonumber \\
&& \langle \delta(\nu), \tilde \delta(\nu)\rangle_{\rm KG}=-2\pi \mu_{r}\delta \, .
\end{eqnarray}
The functions $\alpha(-p)$, $\beta(-p)$ given in (\ref{alphap}), (\ref{betap}) 
are additional Bogolyubov coefficients.
Substituting the expansion (\ref{outfromin}) into the second identity in (\ref{innerprs}) gives 
the second pair of relations for Bogolyubov coefficients (\ref{B_rels2}) where the function 
$d(q_{1}, q_{2})$ is 
$$
d(Q_{1},Q_{2}) = \langle \delta(\nu), \tilde \delta(\nu) \rangle_{KG}[\alpha(-q_{1})\beta(q_{2}) - 
\alpha(q_{2})\beta(-q_{1})]
$$
in accordance with (\ref{dqq}). Thus the extra term on the right hand side of $(\ref{B_rels2})$ indeed 
accounts for the blowing up solution $\tilde \delta(\nu)$. 

The following additional relations can be shown to be true 
\begin{eqnarray}\label{add_rels}
&& \int\limits_{0}^{\infty}\!\! dQ\, [\alpha_{P}^{*}(Q)\beta(q) + \beta^{*}_{P}(Q)\beta(-q)]=0 \, , 
\nonumber \\
&& \int\limits_{0}^{\infty}\!\! dQ\, [\alpha_{P}^{*}(Q)\alpha(q) + \beta^{*}_{P}(Q)\alpha(-q)]=0 \, , 
\nonumber \\
&& \int\limits_{0}^{\infty}\!\! dQ\, [\alpha(-q)\beta(q)-\alpha(q)\beta(-q)]= (\tilde \mu)^{-1}\, .
\end{eqnarray} 
Here and elsewhere $\tilde \mu$ stands for the combination
\begin{equation}
\tilde \mu = 2\pi \mu_{r}\delta \, . 
\end{equation}

The expressions for the modes   ${\cal S}^{\rm in}$ in terms of the modes 
${\cal S}^{\rm out}$  
can be found using the inner products (\ref{alpbeta}), (\ref{innerprs}) and formula  (\ref{outfromin}). 
We derive 
\begin{eqnarray}\label{infromout}
&& \Phi_{-P}(\nu) = \int\limits_{0}^{\infty}\! dQ 
( \beta_{P}(Q)\Psi_{Q}(\nu) + 
 \alpha_{P}(Q) \Psi_{-Q}(\nu) ) \, , \nonumber \\
&&  \delta(\nu)= \tilde \mu \int\limits_{0}^{\infty} \!\!dP[ \alpha(-p)\Psi_{P}(\nu)  - 
\alpha(p)\Psi_{-P}(\nu)     ]\, , \nonumber \\
&& \tilde  \delta(\nu)=
\tilde \mu \int\limits_{0}^{\infty} \!\!dP[ \beta(p)\Psi_{-P}(\nu) - \beta(-p)\Psi_{P}(\nu)      ]\, .
\end{eqnarray}

The expressions (\ref{chi}), (\ref{outfromin}), (\ref{infromout}) define  
(inverse to each other) Bogolyubov transformations between the 
two sets of modes: ${\cal S}^{\rm in}$ and  ${\cal S}^{\rm out}$. 
The corresponding Bogolyubov coefficients are all expressed via the functions  
$\alpha_{P}(Q), \beta_{P}(Q), \alpha(p), \beta(q)$, 
and satisfy the standard unitarity relations. 

The inner product matrix for the basis ${\cal S}^{\rm in}$ can be completely diagonalized 
by introducing 
$$
\delta_{+}=(\tilde \delta(\nu) + \delta(\nu))(2\tilde \mu)^{-1/2}\, , \quad 
\delta_{-}=(\tilde \delta(\nu) - \delta(\nu))(2\tilde \mu)^{-1/2} \, . 
$$
The modes $\delta_{+}$, $\delta_{-}$ are complex conjugates of each other. 
One has 
$$
\langle \delta_{+}, \delta_{+}\rangle_{\rm KG}=-1\, , \quad 
\langle \delta_{-}, \delta_{-}\rangle_{\rm KG}=1\, , \quad 
\langle \delta_{+}, \delta_{-}\rangle_{\rm KG}=0\, . 
$$

Let us introduce a linear operator $ A$ that corresponds 
to a block of the  Bogolyubov transformation (\ref{outfromin}), (\ref{infromout}) 
 that maps the positive frequency 
modes $\Psi_{-P}(\nu)\, , \enspace P \ge 0$ to the modes 
$\Phi_{-P}(\nu)\, ,  P > 0\, , \, \, \delta_{-}  $. It can be written as a block matrix 
\begin{equation}
A= \left[ 
\begin{array}{c}
\alpha_{P}(Q)\\
\alpha(q) + \beta(q) 
\end{array}
\right]\, . 
\end{equation}
Here $Q$ is a column's label.
Similarly the block mapping the negative frequency out states $\Psi_{-P}(\nu)$, $  P \le 0$ to the modes 
$\Phi_{-P}(\nu)\, ,  P > 0\, , \, \, \delta_{-}  $ is given by a block matrix 
 \begin{equation}
B= \left[ 
\begin{array}{c}
\beta_{P}(Q)\\
\alpha(-q) + \beta(-q) 
\end{array}
\right]\, .
\end{equation}
In terms of the operators $A$, $B$ the first relation in (\ref{B_rels2})  can be  rewritten as
\begin{equation}
A^{\dagger}A = I + BB^{\dagger} \, . 
\end{equation}
This implies that the operator $A$ has the bounded inverse (e.g. see \cite{Ber} Chapter 2, section 4.2).
The operator $A^{-1}$ has a block structure
\begin{equation}
A^{-1}= \Bigl[ \gamma_{Q}(P)\enspace
\gamma(q) 
\Bigr]
\end{equation}
where rows are labeled by $Q$\, . In terms of the block entries we obtain the following set 
of relations which will be used later
\begin{eqnarray}\label{inverse_rels}
&& \int\limits_{0}^{\infty}\!\!dP\, \gamma_{Q_{1}}(P)\alpha_{P}(Q_{2}) = \delta(Q_{1}-Q_{2}) 
-\gamma(q_{1})(\alpha(q_{2}) + \beta(q_{2})) \, , \nonumber \\
&&\int\limits_{0}^{\infty}\!\!dQ\, \alpha_{P_{1}}(Q)\gamma_{Q}(P_{2})  = \delta(P_{1}-P_{2})\, , \nonumber \\
&& \int\limits_{0}^{\infty}\!\!dQ\, \alpha_{P}(Q)\gamma(q) = 0 \, , 
\quad \int\limits_{0}^{\infty}\!\!dQ\, (\alpha(q) + \beta(q))\gamma_{Q}(P) = 0 \, , \nonumber \\
&& \int\limits_{0}^{\infty}\!\!dQ\, (\alpha(q) + \beta(q)) \gamma(q) = 1 \, . 
\end{eqnarray}

\section{Secondary quantization}

\setcounter{equation}{0}
Let us discuss now what our findings imply for a secondary  quantization of the 
wave equations (\ref{diffeqs}). A quantum field $\hat \Phi$ can be decomposed in two ways as 
\begin{eqnarray}\label{modes3}
&& \hat \Phi = \int\limits_{0}^{\infty}\! \! dP \, [\Phi_{-P}(\nu) a^{\rm in}_{P}  + 
\Phi_{P}(\nu) a^{\rm in \dagger}_{P}] + 
\tilde \mu^{-1/2} (\hat q \tilde \delta(\nu) -i \hat p  \delta(\nu) ) \, , \nonumber \\ 
&& \hat \Phi = \int\limits_{0}^{\infty}\! \! dP \, [\Psi_{-P}(\nu) a^{\rm out}_{P}  + 
\Psi_{P}(\nu) a^{\rm out \dagger}_{P}] 
\end{eqnarray}
where the creation and annihilation operators satisfy the canonical commutation relations 
\begin{eqnarray}
&& [a^{\rm in}_{P_{1}}, a^{\rm in \dagger }_{P_{2}}] =\delta(P_{1}-P_{2}) \, , 
\quad [a^{\rm in}_{P_{1}}, a^{\rm in  }_{P_{2}}] =0\nonumber \\
&& [a^{\rm out}_{P_{1}},a^{\rm out \dagger}_{P_{2}}] = \delta(P_{1}-P_{2})\, , 
\quad [a^{\rm out}_{P_{1}}, a^{\rm out }_{P_{2}}] =0 \, .
\end{eqnarray}
The operator modes $\hat q$ and $\hat p$ are hermitean and the commutation relations involving these 
operators  can be 
easily determined. Recall that the symplectic form on the space of classical solutions 
can be defined as
\begin{equation}\label{SF}
\Omega(\Phi_{1}, \Phi_{2}) = \langle \Phi_{1}, \Phi_{2}^{*} \rangle_{\rm KG} \, . 
\end{equation}  
Following the rules of second quantization with this symplectic form we obtain from (\ref{innerprs}), (\ref{outfromin})
\begin{equation}
[\hat q, \hat p]=i\, , \quad       [\hat p, a^{\rm in}_{P}]=0 \, , \quad [\hat q, a^{\rm in}_{P}]=0\, . 
\end{equation}


Substituting (\ref{outfromin}) into the second equation in (\ref{modes3}) we obtain
\begin{eqnarray}\label{aa1}
&&a^{\rm in}_{P}=  \int\limits_{\epsilon}^{\infty}\! \! dQ\, 
[\alpha_{P}^{*}(Q)a_{Q}^{\rm out} - \beta^{*}_{P}(Q)a_{Q}^{\rm out \dagger}   ]\, ,  \nonumber \\
&&\hat p  = i\tilde \mu^{1/2}  \int\limits_{0}^{\infty}\! \! dP\, 
[ \beta(-p) a^{\rm out}_{P}  
 +  \beta(p) a^{\rm out \dagger}_{P}     ] \, ,  \nonumber \\
&& \hat q = \tilde \mu^{1/2} \int\limits_{0}^{\infty}\! \! dP\, [\alpha(-p) a^{\rm out}_{P}  
 +  \alpha(p) a^{\rm out \dagger}_{P}   ]\, . 
\end{eqnarray}

The inverse set of linear relations is found by substituting (\ref{infromout}) into the first
 equation in (\ref{modes3}).
 We have
 \begin{equation}
 a_{P}^{\rm out}=\int\limits_{0}^{\infty}\!\!dQ\, [\alpha_{Q}(P)a^{\rm in}_{Q} + \beta^{*}_{Q}(P)a_{Q}^{\rm in \dagger}] 
 + \tilde \mu^{1/2}(\beta(p)\hat q + i \alpha(p)\hat p)\, .
  \end{equation}

Note that the modes $a^{\rm in}_{P}$, $a^{\rm out}_{P}$ each describes a harmonic oscillator with frequency 
$P$ in the respective asymptotic regions.
It is straightforward to define the \out vacuum $|0\rangle_{\rm out}$ as a state 
annihilated by all $a^{\rm out}_{P}$ operators. To define the \inn vacuum $|0\rangle_{\rm in}$ 
 it is  natural to require 
$a^{\rm in}_{P}|0\rangle_{in}=0$ for all $P>0$.  In addition to this 
we need to specify how the  operators $\hat q$,  $\hat p$ act on $|0\rangle_{\rm in}$. 
The modes $\hat q$,  $\hat p$ are canonical variables asymptotically describing a quantized upside-down harmonic 
oscillator. The last one is classically described by the equation 
\begin{equation}\label{udosc}
\partial_{t}^{2}\eta = \delta^{2}\eta
\end{equation}
which is the asymptotic  equation of motion for the mode $\eta$ in the infinite past. 
The corresponding quantum Hamiltonian is unbounded and there is no natural choice of  the 
incoming state for this system. We will discuss the physics of this  mode 
further in section \ref{vacuum}.  For now we record that there is an essential ambiguity in defining 
$|0\rangle_{\rm in}$. 

However it is not hard to see that there is a large class of pair creation amplitudes 
independent of the choice of initial wavefunction for modes $\hat q$,  $\hat p$. 
Apply both sides of the first equation in (\ref{aa1}) to the \inn vacuum. We obtain  
$$
 \int\limits_{0}^{\infty}\!\!dQ\,  \alpha^{*}_{P}(Q)a_{Q}^{\rm out}|0\rangle_{\rm in} = 
  \int\limits_{0}^{\infty}\!\!dQ \, \beta^{*}_{P}(Q)a_{Q}^{\rm out \dagger} |0\rangle_{\rm in} \, . 
   $$
Using this relation and the canonical commutation relations we obtain 
 \begin{equation}\label{ampl1}
 \int\limits_{0}^{\infty}\!\!dQ_{1}\!\! \int\limits_{0}^{\infty}\!\!dQ_{2} \, \alpha_{P_{1}}^{*}(Q_{1})
  \alpha_{P_{2}}^{*}(Q_{2}) {}_{\rm out} \langle Q_{1},Q_{2}| 0\rangle_{\rm in} = 
  \int\limits_{0}^{\infty}\!\!dQ \, \alpha^{*}_{P_{1}}(Q)\beta^{*}_{P_{2}}(Q)
      \end{equation}
  where ${}_{\rm out} \langle Q_{1},Q_{2}|$ is the bra vector conjugated 
  to\footnote{We chose to normalize our particle states so that their inner products involve only appropriate 
  delta functions. Alternatively one may wish the normalizations to be invariant under the asymptotic Lorentz 
  transformations. In that case one should include the factors $\sqrt{|P|}$ with every creation operator 
  $a^{\rm in\dagger}_{P}$, $a^{\rm out\dagger}_{P}$. All  formulas we obtain can be trivially generalized to include such factors. }  
  $|Q_{1},Q_{2}\rangle_{\rm out} = a_{Q_{1}}^{\rm out \dagger}a_{Q_{1}}^{\rm out \dagger}|0\rangle_{\rm out}$.
Formula (\ref{ampl1}) means that a creation amplitude for two \out particles whose wave functions are of the form 
\begin{equation} \label{wavefsp}
\psi(Q) = \int\limits_{0}^{\infty}\!\! dP\,   \phi(P) \alpha_{P}(Q)
\end{equation}
with any sensible weight function $\phi(P)$,
 are independent of how the action of the modes $\hat q$, $\hat p$ is defined on 
$|0\rangle_{\rm in}$. Such amplitudes  are all expressible via the \out wave functions and 
the Bogolyubov coefficients $\alpha_{P}(Q)$ and $\beta_{P}(Q)$. A natural question arises - how big is the space of 
all such functions? This is measured by the dimension of the cokernel of  the operator defined in  (\ref{wavefsp}). 
The first relation in (\ref{inverse_rels}) implies that the cokernel of operator (\ref{wavefsp}) has at most 
dimension one. Thus almost all pair creation amplitudes are independent of the details of the initial 
state of the tachyonic mode $\eta$. In the second quantized approach these amplitudes are expressed via 
Bogolyubov coefficients   (\ref{ampl1}).
In the next section we will show that pair creation amplitudes (\ref{ampl1})  can be obtained by 
computing an appropriate string theory two-point function.


\section{String theory two point function} \label{2ptsec}
\setcounter{equation}{0}
As was said  in the introduction  it is natural to expect 
 that string theory two point functions in time-dependent backgrounds 
are related to pair creation amplitudes which are expressible via Bogolyubov coefficients (see also 
\cite{GS}). 
We would like to check this relationship for the set of amplitudes 
described in the previous section. We will show that for a properly defined string two point function 
the following relation holds 
\begin{eqnarray}\label{shouldbe}
\frac{1}{2}\langle  \Phi_{-P_{1}}(\nu)\Phi_{-P_{2}}(\nu)  \rangle_{\rm str} &= &
\int\limits_{0}^{\infty}\!\!dQ_{1}\!\! \int\limits_{0}^{\infty}\!\!dQ_{2} \, \alpha_{P_{1}}(Q_{1})
  \alpha_{P_{2}}(Q_{2}) {}_{\rm in} \langle 0| Q_{1},Q_{2} \rangle_{\rm out} \nonumber \\
  &=& \int\limits_{0}^{\infty}\!\!dQ\, 
 \alpha_{P_{1}}(Q)\beta_{P_{2}}(Q)    \, . 
\end{eqnarray} 

Defining a two point amplitude in string theory involves 
 fixing the $SL(2, {\mathbb R})$ modular symmetry. Fixing the positions of  two
 vertex operators leaves out a subgroup of infinite volume. Dividing over this infinite volume typically yields a
 vanishing amplitude. 
  In some cases  a finite quantity can be obtained 
 by cancelling the infinite volume of the modular group against the infinte factor $\delta(0)$ arising from the target space 
zero mode integration \cite{Seiberg}, \cite{GM}. 
This cancellation is relatively well understood in noncritical string theory but is also believed to happen in other 
models for example for strings propagating in $AdS_{3}$ \cite{morecomments}. As there is no general lore we offer 
only a few comments on this issue which hopefully clarify the situation to some extent. 

While the two infinite factors are in general  unrelated, in noncritical string theory dilatations involve translations 
of the Liouville field because of the background charge. At the technical level one derives two point functions 
in noncritical string theory by starting with a three point amplitude which is free of divergences 
and using the ground ring structure that relates it to two point functions  \cite{DFrKut}, \cite{Kostov}. 
Breaking of target space translation invariance is crucial in this approach because the three point function 
does not have a momentum conservation delta function. In the example at hand we do not know if there is some algebraic structure 
similar to the ground ring so we proceed in a somewhat empirical fashion. We first investigate the CFT two point function. 
The formal expression turns out to be divergent. We investigate the nature of the divergences by regularizing 
the amplitude and find that the divergences come from contributions of the on-shell states of the undeformed  theory 
describing FZZT branes. We then use the results of \cite{Kostov}  for string boundary two point functions 
for FZZT branes to  cancel  the aforementioned divergences against the 
modular group volume. Having sketched the idea we now give the details.

Using  (\ref{specs})  we find that  the  formal expression for  the CFT two point  function has the following contributions 
$$
\langle  \Phi_{-P_{1}}(\nu), \Phi_{-P_{2}}(\nu)  \rangle_{\rm CFT} = C_{0}(P_{1},P_{2})  + C_{1}(P_{1},P_{2}) 
+ C_{1}(P_{2},P_{1}) + C_{\mu}(P_{1},P_{2}) + C_{\eta}(P_{1},P_{2}) 
$$
where 
$$
C_{0}(P_{1},P_{2})= \xi_{1}\xi_{2}\langle \Phi_{|P_{1}|}^{\delta}e^{-iP_{1}}, 
\Phi_{|P_{2}|}^{\delta}e^{-iP_{2}X_{0}}\rangle_{\rm CFT} \, , 
$$
$$
 C_{1}(P_{1},P_{2})= \xi_{1}\xi_{2}\int\limits_{-\infty}^{+\infty}\!\!d\omega \int\limits_{0}^{\infty}\!\!dQ 
 \frac{f(Q)d_{-P_{2}}(\omega)}{(\omega + i\epsilon)^{2}- Q^{2}}  
 \langle \Phi_{|P_{1}|}^{\delta}e^{-iP_{1}X_{0}}, 
\Phi_{Q}^{\delta}e^{-i\omega X_{0}}\rangle_{\rm CFT} \, , 
$$
\begin{eqnarray*}
 C_{\mu}(P_{1},P_{2})=&&\xi_{1}\xi_{2}\int\limits_{-\infty}^{+\infty}\!\!d\omega_{1} \int\limits_{-\infty}^{+\infty}\!\!d\omega_{2} 
\int\limits_{0}^{\infty}\!\!dQ_{1} \int\limits_{0}^{\infty}\!\!dQ_{2}
 \frac{f(Q_{1})f(Q_{2}) d_{-P_{1}}(\omega_{1}) d_{-P_{2}}(\omega_{2})}
 {[(\omega_{1} + i\epsilon_{1})^{2}- Q^{2}_{1}][(\omega_{2} + i\epsilon_{2})^{2}- Q^{2}_{2}]}   \nonumber \\
&&  \langle \Phi_{Q_{1}}^{\delta} 
e^{-i\omega_{1} X_{0}}, \Phi_{Q_{2}}^{\delta}e^{-i\omega_{2} X_{0}}\rangle_{\rm CFT} \, , 
\end{eqnarray*}

$$
C_{\eta}(P_{1},P_{2})=4\delta^{2}\xi_{1}\xi_{2}\int\limits_{-\infty}^{+\infty}\!\!d\omega_{1} \int\limits_{-\infty}^{+\infty}\!\!d\omega_{2} 
\frac{ d_{-P_{1}}(\omega_{1}) d_{-P_{2}}(\omega_{2})}{(\delta^{2}+ \omega_{1}^{2})(\delta^{2}+ \omega_{2}^{2}) }
\langle \Phi_{\vartheta}^{\delta} 
e^{-i\omega_{1} X_{0}}, \Phi_{\vartheta}^{\delta}e^{-i\omega_{2} X_{0}}\rangle_{\rm CFT}
$$
Here for brevity we denoted the normalization factors as $\xi_{1}=\xi_{0}(-P_{1})$, $\xi_{2}=\xi_{0}(-P_{2})$ (see (\ref{xi0})). 
We also dropped the coordinate dependence of the two point functions: the notation $\langle {\cal O}_{1}, {\cal O}_{2}\rangle_{\rm CFT}$ 
is used for   two point functions on a unit disc with operators $ {\cal O}_{1}, {\cal O}_{2}$ inserted at  opposite points on the disc boundary.  

Except for the last one all of the above four  expressions contain divergences. 
We will see that these divergences arise from integration over the target space zero modes. 
The first expression can be readily computed using 
(\ref{2p1}) and 
\begin{equation}\label{X02pt}
\langle e^{-iX_{0}\omega_{1}}(x_{1}) e^{-iX_{0}\omega_{2}}(x_{2}) \rangle_{\rm CFT} = 
\delta(\omega_{1}+ \omega_{2})|x_{1}- x_{2}|^{-2\omega_{1}^{2}}
\end{equation}
to obtain 
$$
C_{0}(P_{1},P_{2}) = \xi_{1}\xi_{2}|C_{\delta}(P_{1})|^{2}\delta(|P_{1}|-|P_{2}|)\delta(P_{1}+P_{2}) \, .
$$
This expression contains $\delta(0)$ when $P_{1}$ and $P_{2}$ have opposite signs but this is not the 
case we are interested in for computing the particle production rate. We assume both $P_{1,2}>0$ 
and thus can  set\footnote{This can be done more mathematically 
accurately by first regularising 
this expression, as will be done below, and then removing the regulator.} $C_{0}(P_{1},P_{2}) =0 $.

To exhibit the nature of divergences in $C_{1}(P_{1},P_{2})$ and $C_{\mu}(P_{1},P_{2})$  we 
regularize the above expressions by inserting an extra exponent $e^{-i\sigma X_{0}}$ inside the normal product 
in each 
correlator of the form 
 $\langle \Phi_{Q_{1}}^{\delta} 
e^{-i\omega_{1} X_{0}}, \Phi_{Q_{2}}^{\delta}e^{-i\omega_{2} X_{0}}\rangle_{\rm CFT}$
so that it is replaced by 
$$
\langle \Phi_{Q_{1}}^{\delta} 
e^{-i\omega_{1} X_{0}}, \Phi_{Q_{2}}^{\delta}e^{-i(\omega_{2} + \sigma) X_{0}}\rangle_{\rm CFT}= 
|C_{\delta}(Q_{1})|^{2}\delta(Q_{1}-Q_{2})\delta(\omega_{1}+\omega_{2}+\sigma) \, . 
$$
Substituting this into the above expressions for $C_{1}(P_{1},P_{2})$   we 
obtain 
\begin{equation}
C_{1}^{\sigma} = \xi_{1}\xi_{2}|C_{\delta}(P_{1})|^{2}f(P_{1})
\frac{d(-P_{1}-\sigma)}{(-P_{1}-\sigma +i\epsilon)^{2}-P_{1}^{2}}\, .
\end{equation}
We observe that in the limit $\sigma \to 0$ the divergence comes from 
restricting on the mass shell the expression  $((\omega + i\epsilon)^{2}- Q^{2})^{-1}$. 
The later can   be interpreted as 
the asymptotic free field propagator. The divergence has the form 
\begin{equation}\label{div1}
C_{1}^{\rm div} = \xi_{1}\xi_{2}|C_{\delta}(P_{1})|^{2}f(P_{1})
\frac{d(-P_{1})}{2P_{1}\sigma} \, . 
\end{equation}
After the same  substitution  done for $C_{\mu}(P_{1},P_{2})$ we obtain 
by taking integrals in $Q_{1}$, $\omega_{1}$
\begin{equation}
C_{\mu}^{\sigma} = 2\mu_{r}\xi_{1}\xi_{2}\int\limits_{-\infty}^{+\infty}\!\! d\omega
d_{-P_{1}}(\omega)d_{-P_{2}}(-\omega -\sigma) I^{\sigma}(\omega)
\end{equation}
where 
$$
I^{\sigma}(\omega) = \int\limits_{-\infty}^{+\infty}\!\!
\frac{Q^{2}dQ}{[(\omega + i\epsilon_{1})^{2}-Q^{2}][(\omega + \sigma -i\epsilon_{2})^{2}-Q^{2}](\delta^{2}+Q^{2})}\, .
$$
The integral $I^{\sigma}(\omega)$ can be computed by taking the residues in the complex upper half plane 
at $Q=i\delta$, $Q=\omega + i\epsilon_{1}$, $Q=-\sigma-\omega + i\epsilon_{2}$. We obtain 
$$
I^{\sigma}(\omega)=-\frac{\pi\delta}{(\omega^{2}+\delta^{2})^{2}} - \frac{\pi i}{\sigma(\omega^{2}+\delta^{2})} \, . 
$$
Here the second term which is divergent came from the residues taken at $Q=\omega + i\epsilon_{1}$, 
$Q=-\sigma-\omega + i\epsilon_{2}$. Thus again, as in the case of $C^{\sigma}_{2}$, the divergences come when one of the 
asymptotic propagators is put  on shell. The divergence can be isolated to be 
\begin{equation}\label{div2}
C_{\mu}^{\rm div} = -\frac{2\pi i \mu_{r}\xi_{1}\xi_{2}}{\sigma}\int\limits_{-\infty}^{+\infty}\!\!d\omega 
 \frac{ d_{-P_{1}}(\omega)d_{-P_{2}}(-\omega) }{\omega^{2}+\delta^{2}}  
\end{equation} 

The point of doing the above calculations in detail  was to show that the divergences arise when 
one of the  expressions of the form  $((\omega + i\epsilon)^{2}- Q^{2})^{-1}$ is reduced to one of 
the values $\omega=\pm Q=\pm P_{1,2}$. Since such an expression originates as a term in  (\ref{specs}) 
that stands at $\Phi_{Q}^{\delta}e^{-i\omega X_{0}}$ we see that the corresponding state is 
projected onto an on-shell value. Furthermore the propagator itself contains  on-shell delta functions 
$\delta(Q\pm \omega)$ so that we can interpret the divergence at hand as the one arising in 
a CFT two-point function for two on-shell operators in the undeformed theory: 
$\langle \Phi_{Q_{1}}^{\delta} 
e^{-iQ_{1} X_{0}}, \Phi_{Q_{2}}^{\delta}e^{-iQ_{2} X_{0}}\rangle_{\rm CFT}$.
The last expression formally evaluated contains $\delta(0)$. 

So far we have considered only the CFT two-point function.  
For open strings of the FZZT brane in two dimensional string theory 
a consistent string theory two point function can be extracted from the ground ring relations 
\cite{Kostov}. Formula (D.12) of \cite{Kostov} taken for $b=1$ in our notations 
reads\footnote{To obtain  formula (\ref{Kost}) from  formula (D.12) of \cite{Kostov} one needs to take into account 
different  normalizations given in (2.9) of \cite{Kostov} and the fact that in \cite{Kostov} 
a Euclidean signature time field is considered. The latter  results in an overall  sign change.}  
\begin{equation}\label{Kost}
\langle \Phi_{Q_{1}}^{\delta} 
e^{-iQ_{1} X_{0}}, \Phi_{Q_{2}}^{\delta}e^{-iQ_{2} X_{0}}\rangle_{\rm str} = 2\pi Q_{1} 
|C_{\delta}(Q_{1})|^{2}\delta(Q_{1}-Q_{2}) \, . 
\end{equation}
(It is assumed here that $Q_{1,2}>0$.) 
The appearance of the $Q_{1}$ factor in the above expression was to be expected from relativistic 
invariance in the asymptotic  region. 

The fact that the divergences in a two point function of the time-dependent theory 
arise from  overlaps of pairs of states that are on-shell in the undeformed theory 
implies that we can use formula (\ref{Kost}) to extract finite results from those divergent pieces. 
The string theory two point function can be defined as\footnote{The numerical factor in  
(\ref{str2pt_def})  depends on the particular choice of the regularization parameter $\sigma$.
We fixed  this ambiguity essentially by hand to yield the correct final expression. 
A more satisfactory solution to this problem would  involve identifying uniquely a $\sigma$-regularization 
of the volume of residual  modular group.}
\begin{equation} \label{str2pt_def}
\langle  \Phi_{-P_{1}}(\nu)\Phi_{-P_{2}}(\nu)  \rangle_{\rm str} := 
2\pi i \lim\limits_{\sigma \to 0}\sigma \langle \Phi_{-P_{1}}(\nu), :e^{-i\sigma X_{0}}
\left|\frac{\partial}{\partial X_{0}}\right|\Phi_{-P_{2}}(\nu):\rangle_{\rm CFT} 
\end{equation}
where  $|\frac{\partial}{\partial X_{0}}|$ stands for an operator formally defined as 
$$
:\left|\frac{\partial}{\partial X_{0}}\right|e^{-i\omega X_{0}}: = |\omega|:e^{-i\omega X_{0}}: \, . 
$$
With this prescription applied to the divergent parts (\ref{div1}), (\ref{div2}) we readily obtain 
an expression that can be recognized as the expression in the right hand side 
of (\ref{shouldbe}) multiplied by a factor of 2, that is
\begin{equation} \label{result}
\langle  \Phi_{-P_{1}}(\nu)\Phi_{-P_{2}}(\nu)  \rangle_{\rm str} = 
\int\limits_{0}^{\infty}\!\!dQ
 [\alpha_{P_{1}}(Q)\beta_{P_{2}}(Q) + \alpha_{P_{2}}(Q)\beta_{P_{1}}(Q)] = 
 2\int\limits_{0}^{\infty}\!\!dQ \alpha_{P_{1}}(Q)\beta_{P_{2}}(Q) \,  
\end{equation}
where in the last step we used the second relation in (\ref{B_rels}).
Formula (\ref{result})  is the main result of this section. It means that for the class of outgoing particles 
with wavefunctions (\ref{wavefsp}) the pair creation amplitudes computed via Bogolyubov coefficients 
(\ref{B_coefs}) 
coincide with a suitably defined string theory two-point function.


\section{String theory three point function} \label{3ptsec}
\setcounter{equation}{0}
In this section we will compute the string three point function of the operators $\Phi_{-P}(\nu)$. 
This amplitude does not contain any divergences and the computation is straightforward. It boils 
down to computing the CFT three point function which is then stripped of its dependence on the 
insertion points. Thus in the following we will  suppress the insertion points. 

We start by substituting  expansions (\ref{newops}), (\ref{specs})  into the three point function
\begin{equation}
\langle \Phi_{-P_{1}} \Phi_{-P_{2}}\Phi_{-P_{3}} \rangle_{\rm CFT}
\end{equation}
and using (\ref{3ptL}), (\ref{X02pt}). This yields the following expression
\begin{equation}
\langle \Phi_{-P_{1}} \Phi_{-P_{2}}\Phi_{-P_{3}} \rangle_{\rm str} = 2\pi\!\!
 \int\limits_{-\infty}^{\infty}\!\!dt\, 
h(t, -P_{1})h(t, -P_{2})h(t, -P_{3}) 
\end{equation}
where the function $h(t, -P)$ is given in (\ref{hsol}).
The above integral can be conveniently computed by using the Fourier transform of the integrand. 
The integrand up to an exponential factor $\exp(-it(P_{1}+P_{2}+P_{3}))$ is 
 a polynomial in functions $W(t)$. The Fourier transforms of powers $W^{n}(t)$ can be readily 
 computed using the differential equation 
 \begin{equation}
 W^{2} = \delta W - \partial_{t}W \, . 
 \end{equation}
This differential equation implies the recurrence relation for the Fourier transforms $\widehat{W^{n}}(\tilde \omega)$
\begin{equation}
\widehat{W^{n+1}}(\tilde \omega)=  \delta (1 + \frac{i\tilde \omega}{n})\widehat{W^{n}}(\tilde \omega)
\end{equation}
that can be solved as 
\begin{equation}\label{Wn}
\widehat{W^{n}}(\tilde \omega)=\delta^{n}\frac{\Gamma(n+i\tilde \omega)}{\Gamma(1+i\tilde \omega)(n-1)!}\hat W(\tilde \omega) \, .
\end{equation}
This implies that the amplitude has the following form 
\begin{equation}
\langle \Phi_{-P_{1}} \Phi_{-P_{2}}\Phi_{-P_{3}} \rangle_{\rm str} = 
\frac{\xi_{0}(-p_{1}) \xi_{0}(-p_{2}) \xi_{0}(-p_{3})P(p_{1},p_{2},p_{3})}{p_{1}p_{2}p_{3}(p_{1}+i)(p_{2}+i)(p_{3}+i)}\hat W(p_{1}+p_{2}+p_{3})
\end{equation}
where $P(p_{1},p_{2},p_{3})$ is a polynomial. The last one can be computed using (\ref{Wn}). Using Maple we arrive at the following compact 
looking result 
\begin{equation} \label{3ptfn}
\langle \Phi_{-P_{1}} \Phi_{-P_{2}}\Phi_{-P_{3}} \rangle_{\rm str}= \frac{2\pi}{15}F(p_{1})F(p_{2})F(p_{3})
\frac{(\Pi(p_{1}) + \Pi(p_{2}) + \Pi(p_{3}))}{\sinh[\pi(p_{1}+p_{2} + p_{3})]}
\end{equation}
where
\begin{equation}
F(p_{i})= \frac{\nu^{ip_{i}}}{ \pi \sqrt{\mu_{r}\delta p_{i}} (1 + p_{i}^{2})}\, , 
\end{equation}
\begin{equation}
\Pi(p) = 2p + 5p^{3} + 3p^{5} \, .
\end{equation}
We conjecture that this expression, being integrated with wave function factors $\phi_{i}(p_{i})$, $i=1,2,3$, 
gives a triplet creation amplitude due to string interaction. The outgoing states in this amplitude have 
 wavefunctions of the form (\ref{wavefsp}).


\section{Choice of the \inn vacuum}\label{vacuum}
\setcounter{equation}{0}
In this section we will discuss  how one can specify  completely a reasonable set of \inn vacua and 
will find the corresponding expressions in terms of the \out states. 
To complete the definition of $|0\rangle_{\rm in}$ we need to specify the initial quantum state of the 
$\eta$ mode described by the upside-down harmonic oscillator (\ref{udosc}) with symplectic form (\ref{SF}). 

Note that the value of $\nu$ can be offset by a suitable time translation so from now on 
we will set it equal to $1$.
The canonically conjugated momentum to the quantized coordinate $\hat \eta$ is  
$$
\hat \pi =\frac{\tilde \mu}{4\delta^{2}}\widehat{\partial_{t} \eta}\,  
$$ 
and the Hamiltonian reads
\begin{equation} \label{Hamm}
\hat H = \frac{2\delta^{2}}{\tilde \mu}\hat \pi^{2} -\frac{\tilde \mu}{8}\hat \eta^{2} \, . 
\end{equation}

The canonical pair $\hat q$ and $\hat p$ is related to $\hat \eta$, $\pi$ as 
\begin{eqnarray}
&& \hat q = \frac{1}{2}\tilde \mu^{1/2}\hat \eta - 2\delta \tilde \mu^{-1/2}\hat \pi \, , \nonumber \\
&& \hat p = \frac{\tilde \mu^{1/2}}{4\delta}\hat \eta +  \tilde \mu^{-1/2}\hat \pi \, .
\end{eqnarray}

Since the Hamiltonian (\ref{Hamm}) is unbounded from below there is no vacuum state for this system. 
On the other hand any reasonable initial state for the decaying FZZT brane should have the tachyonic 
mode localized around the  zero value. We thus require that 
\begin{equation}\label{localization}
{}_{\rm in}\langle 0|\hat \eta^{2}|0\rangle_{\rm in} = a \, . 
\end{equation} 
 Since the value of the $\eta$ variable in the new vacuum is $u_{*}=2\delta$ it is reasonable to 
 require $a\ll 4\delta^{2}$.   
 Among all states satisfying the constraint (\ref{localization}) one can find 
  the state for the $\hat \eta$, $\hat \pi$ system that minimizes the expectation value of the 
   energy  (\ref{Hamm}) \cite{WW}, \cite{GuthPi}. The result is a Gaussian wave function
   \begin{equation} \label{istate}
   \psi_{0}(\eta)= (2a\pi)^{-1/4}e^{-\eta^{2}/4a} \, . 
   \end{equation}  
 The time evolution of this wave function is \cite{GuthPi}
 \begin{eqnarray}
 && \psi(\eta, t)= A(t)e^{-\eta^{2}B(t)}\, , \nonumber \\
 && B(t)=\frac{\tilde \mu}{8\delta}\tan(\phi - i\delta t)\, , \quad 
 A(t)=(2\pi)^{-1/4}[b\cos(\phi-i\delta t)]^{-1/2} \, \nonumber \\
 && \phi= \arctan\Bigl(\frac{2\delta}{\tilde \mu a}\Bigr)\, , \quad 
 b=(\sin(2\phi))^{-1/2}(4\delta/\tilde \mu)^{1/4} \, . 
 \end{eqnarray}  
  For any value of $a$ the wave packet rapidly spreads. For small values of $a$ this happens due to 
  the uncertainty principle while for large values due to the unboundedness of the potential.
  For large times the speed of the spread is exponential, proportional to $e^{\delta t}$.
  The $\eta$ degree of freedom however is described by an upside down oscillator only asymptotically, 
  at  some point the interaction effects  become significant with the energy  of the $\eta$ 
  degree of freedom being lost into  radiation.

  The initial wave function (\ref{istate}) satisfies 
  \begin{equation}\label{supp_cond}
  \hat \pi \psi_{0}(\eta) = \frac{i}{2a}\eta \psi_{0}(\eta) \, .  
  \end{equation}
  In terms of the operators $\hat q$, $\hat p$ this condition reads 
  \begin{equation}\label{condd}
\hat p \psi_{0} = C_{a}\hat q \psi_{0}\, , \quad 
C_{a}=\frac{(1 + i\frac{2\delta}{\tilde \mu a})}{2\delta(1 - i\frac{2\delta}{\tilde \mu a})}\, . 
\end{equation}
We  add the condition $\hat p |0\rangle_{\rm in} = C_{a}\hat q |0\rangle_{\rm in}$ to 
the conditions $a_{P}^{\rm in}|0\rangle_{\rm in}=0$ characterizing the \inn state. 
 These equations  have a unique solution in terms of the \out oscillators. 
Using (\ref{inverse_rels}), (\ref{add_rels}) one finds that 
a solution to equations $a_{P}^{\rm in}|0\rangle_{\rm in}=0$ has the following general form 
\begin{equation}
|C\rangle = F[\hat q]\, \hat G |0\rangle_{\rm out} 
\end{equation}
where 
\begin{equation}
 \hat G= \exp\Bigl[ \frac{1}{2} \int\limits_{0}^{\infty} 
 \!\! \int\limits_{0}^{\infty}\!\!
 dQ_{1}dQ_{2}\, [\int\limits_{0}^{\infty}\!\!dP \gamma^{*}_{Q_{1}}(P)\beta^{*}_{P}(Q_{2}) 
 + \gamma^{*}(q_{1})(\beta(q_{2})-\alpha(q_{2}))]
 \, a^{\rm out\dagger}_{Q_{1}} 
 a^{\rm out\dagger}_{Q_{2}}     \Bigr] \, . 
\end{equation}
One further finds that  
\begin{equation}
\hat p \hat G |0\rangle_{\rm out} = i \hat q \hat G |0\rangle_{\rm out} \, . 
\end{equation}
Equation (\ref{condd}) fixes the function $F$ so that the state
\begin{equation}\label{astates}
|a\rangle_{\rm in } \equiv \exp\Bigl[ \frac{1 + iC_{a}}{2}\hat q^{2} \Bigr] \hat G |0\rangle_{\rm out} 
\end{equation}
solves the conditions  $a_{P}^{\rm in}|a\rangle_{\rm in}=0$ and 
$\hat p |a\rangle_{\rm in} = C_{a}\hat q |a\rangle_{\rm in}$. It can be used 
to compute generic pair creation amplitudes ${}_{\rm out}\langle Q_{1}Q_{2}|a\rangle_{\rm in}$. 
It would be interesting to find out whether there is a prescription for computing such 
amplitudes via a suitable string theoretic  two point function. 
Using (\ref{infromout}), (\ref{inverse_rels}) one finds that 
\begin{equation}
\int\limits_{0}^{\infty}\!\!dP \, \gamma_{Q}(P) \Phi_{-P}(\nu)  + \sqrt{2/\tilde \mu}\gamma(q)\delta_{-} = \Psi_{-Q}(\nu) + \dots
\end{equation}
where the dots stand for out-states of positive frequency. This combination is a natural candidate for  a vertex operator whose 
string theory two point function gives  he ${}_{\rm out}\langle Q_{1}Q_{2}|a\rangle_{\rm in}$ 
amplitude.  The difficulty in defining a two point function for these operators is in defining it for terms involving  $\tilde \delta(\nu)$. One could 
imagine that a prescription defining the two-point function for exponentially blowing up operators like $\tilde \delta(\nu)$
may involve a parameter $a$ that could be identified with the $a$ present in the definition of $|a\rangle_{\rm in}$. 
We leave the detailed investigation of this question for future work.

\section{Summary and further directions}
In this paper we computed to leading order in $\delta$  string vertex operators for the time-dependent model of 
\cite{2}.
The expressions giving time dependent vertex operators for string states asymptoting to  \inn and \out states 
are given in (\ref{series_exps}), (\ref{specs}), (\ref{tsol1}), (\ref{tsol2}).
Several special solutions at zero momentum were identified (see (\ref{delta_expl}), 
(\ref{Phi0_expl}), (\ref{tildedelta})). 
  We defined bases of \inn and \out scattering states corresponding to scaling operators at the associated UV and IR 
fixed points (see formulas (\ref{in_sts}), (\ref{out_sts})). The complete set of Bogolyubov 
coefficients was obtained in (\ref{chi}), (\ref{outfromin}), (\ref{infromout}) and  unitarity relations between them checked. 
 
We further discussed the second quantization of this system and identified 
a codimension one subspace of out-going wave functions for which the pair creation amplitudes are independent 
of the ambiguity in defining the initial state of the tachyonic mode. We then showed in section  8 that 
this set of amplitudes can be obtained in the first quantized framework by computing the appropriate
 string theoretic two-point functions.  As we discussed 
 in section 8 the main difficulty with computing string two point functions in general is in the need to define a 
 ratio of two infinite factors. Our computation proceeded in a somewhat ad hoc manner, utilizing 
 the known results for  two point functions in noncritical string theories \cite{Seiberg}, \cite{Kostov}. 
 It is clear that a deeper understanding of string theory two point functions and a general method for their computation 
 is needed.  For the same codimension one subspace we computed a string three point functions (\ref{3ptfn}) and 
 conjectured that it gives a triplet creation amplitude due to string interaction. 
 A possible test of this conjecture could come from an open  string field theory description of the model. 
 One could envision a Das-Jevicki type \cite{DJ} theory with a fundamental cubic vertex. The triplet creation amplitude 
 would be then analogous to the one computed in \cite{BF} for a scalar $\phi^{3}$ theory in an expanding universe.   

In the main body of the paper we discussed two special  features  related to the  tachyonic nature of the onset of the time-dependent process at hand. The first feature is 
the need to deal with solutions exponentially blowing up in the far past while the second, related feature, is   the ambiguity in defining 
the quantum initial state of the system (the absence of Fock vacuum). 
We suspect that both of these peculiarities    generically take place in tachyon decay processes.
We discussed the second feature in some detail in section 10 where, following the ideas 
of \cite{GuthPi}, \cite{WW}, we constructed a family of physically reasonable 
initial states (\ref{astates}) in the second quantized oscillator state space.

As for the presence of exponentially blowing up vertex operators in the scattering spectrum 
consider    a string background whose spatial CFT has  relevant operators $\Phi_{i}$. 
Following \cite{FHL} one can 
consider a time-dependent CFT perturbed by a marginal operator of the form $\lambda \Phi_{i}e^{X_{0}p_{i}}$ where 
$\lambda$ is a coupling constant and $p_{i}>0$. Such a background will have an infinitesimal deformation corresponding 
to  varying $\lambda$. In the far past the corresponding vertex operator  vanishes while generically it will not vanish in the far future 
where it is described by  some superposition of outgoing scattering states. Assume that the time-dependent CFT  
has a conserved inner product (\ref{inpr_str}).  We then see that the only way to reconcile the existence of vertex operators 
vanishing in the far past but not in the far future with the conservation of  inner product is to admit the existence of solutions blowing up in the far past. A vertex operator asymptoting to positive frequency \out states will generically 
blow up in the far past. Thus computing generic amplitudes will require making sense of correlators involving 
vertex operators that exhibit such blow up behavior. There may be many prescriptions for making sense of such correlators. 
One would need then to find a rule for matching those prescriptions with initial wave functions for tachyonic modes 
so that the amplitudes computed in the first quantized (CFT) formalism are equal to the corresponding amplitudes 
defined in the second quantization. 
We plan to come back to this issue for the model studied in this paper in future work. 
This problem may test the limits of the first quantized formalism for time-dependent problems. It is worth 
mentioning in this context a problem of UV divergence in the number of emitted  closed string particles 
radiating  from decaying D-branes \cite{LLM}.   
It was shown that the problem that is present in the first quantized approach gets cured for decaying D0 branes 
in two dimensional string theory in the second 
quantized formalism \cite{KMS}. The divergence disappears 
 when one includes in consideration the initial wave function of the D0 brane.

  For the model considered in this paper it would be also interesting to study one-loop amplitudes and the first order  backreaction effects due to radiation of open and closed strings. We leave these questions for future work.       

\setcounter{equation}{0}



\appendix
\renewcommand{\theequation}{\Alph{section}.\arabic{equation}}
\setcounter{equation}{0}
\section{Relations for Bogolyubov coefficients}
Checking relations (\ref{B_rels}) boils down to proving the following identity 
\begin{equation}\label{Aident}
\int\limits_{-\infty}^{+\infty}\! dq\,  \frac{q^{3}}{\sinh[\pi(q-p_{1}-i\epsilon)]
\sinh[\pi(q-p_{2} + i\epsilon')]  } = 
\frac{p_{1}^{2}(p_{1}-i)^{2} - p_{2}^{2}(p_{2}+i)^{2}  }{2\sinh[\pi(p_{2}-p_{1}-i\epsilon)]} 
\end{equation} 
where $p_{1}$ and $p_{2}$ are any two real numbers.  To compute the integral above 
we first compute a generating function 
\begin{equation}
I_{p_{1},p_{2}}(t)=\int\limits_{-\infty}^{+\infty}\! dq\,  \frac{e^{iqt}}{\sinh[\pi(q-p_{1}-i\epsilon)]
\sinh[\pi(q-p_{2} + i\epsilon')]  } \, . 
\end{equation}
It can be evaluated by summing the appropriate residues in the $q$-complex plane 
\begin{equation}
I_{p_{1},p_{2}}(t)=\frac{2i}{\sinh[\pi(p_2 - p_1 -i\epsilon)]}\Bigl(\frac{e^{itp_{2}}e^{-t} - e^{itp_{1}}}{1-e^{-t}}    \Bigr) \, . 
\end{equation}
The right hand side of  identity (\ref{Aident}) can be evaluated now by computing $i[\partial^{3}_{t}I_{p_{1},p_{2}}(t)]_{t=0}$. 

The second pair of relations between Bogolyubov coefficients (\ref{B_rels2}) follows from 
the following identity
\begin{eqnarray}\label{integral2}
&& \mbox{ P.V.}\int\limits_{-\infty}^{+\infty}\!\! dp
\frac{1}{p(1+p^{2})^{2}\sinh[\pi(p-q_{1}-i\epsilon)]\sinh[\pi(p-q_{2}+i\epsilon')]}= \nonumber \\
&& \frac{1}{2\sinh[\pi(q_2-q_1-i\epsilon)]}\Bigl(\frac{1}{q_{2}^{2}(q_2+i)^{2}} -  
 \frac{1}{q_{1}^{2}(q_1-i)^{2}} \Bigr) + \delta(q_{1}, q_{2}) 
 \end{eqnarray}
 where 
 \begin{equation} \label{deltaqq}
 \delta(q_{1}, q_{2}) = -\frac{\pi^{2}}{2}\Bigl(\frac{\cosh(\pi q_{1})}{\sinh^{2}[\pi(q_1 + i\epsilon)]\sinh[\pi(q_2-i\epsilon')]} 
  +  \frac{\cosh(\pi q_{2})}{\sinh[\pi(q_1 + i\epsilon)]\sinh^{2}[\pi(q_2-i\epsilon')]}  \Bigr)\, .
  \end{equation}
This identity can be obtained by taking the integral via summing over the residues in a complex half plane.
Substituting the explicit expressions (\ref{B_coefs}) into the left hand side of (\ref{B_rels2})  
and using (\ref{integral2}) we obtain the right hand side of (\ref{B_rels2}) with 
\begin{equation}\label{dqq}
d(Q_{1},Q_{2}) = -\frac{4}{\delta}|q_{1}q_{2}|^{3/2}\nu^{i(q_{1}-q_{2})}
\left(\frac{q_{1}-i}{q_{1}+i}  \right)\left(\frac{q_{2}+i}{q_{2}- i}  \right) \delta(q_{1}, q_{2}) 
\end{equation}
that can be equivalently written as  
$$
d(Q_{1},Q_{2}) = \langle \delta(\nu), \tilde \delta(\nu) \rangle_{KG}[\alpha(-q_{1})\beta(q_{2}) - 
\alpha(q_{2})\beta(-q_{1})] \, . 
$$


\begin{thebibliography}{99}

\bibitem{1} J. Teschner, {\it On Boundary Perturbations in Liouville Theory and Brane 
Dynamics in Noncritical String Theories}, JHEP 0404 (2004) 023; hep-th/0308140.


\bibitem{2} K. Graham, A. Konechny, J. Teschner, {\it On the time-dependent description 
for the decay of unstable D-branes}, JHEP 0702 (2007) 011; hep-th/0608003.

\bibitem{BPZ} A. Belavin, A. Polyakov and  A. Zamolodchikov, {\it Infinite conformal symmetry in 
two-dimensional quantum field theory}, Nucl. Phys. {\bf B241}, p. 333 (1984).


\bibitem{TeschnerL} J. Teschner, {\it Liouville Theory Revisited}, Class. and Quant. Grav. {\bf 18}
 (2001) R153; hep-th/0104158. 

\bibitem{FZZ} V. Fateev, A. B. Zamolodchikov and A. B. Zamolodchikov, {\it Boundary Liouville 
field theory. I:boundary state and boundary two-point function}, hep-th/0001012.

\bibitem{T} J. Teschner, {\it Remarks on Liouville theory with boundary}, hep-th/0009138. 

\bibitem{PT} B. Ponsot and  J. Teschner, 
{\it Boundary Liouville Field Theory: Boundary Three Point Function }, Nucl. Phys. B622 (2002) 309-327; 
 hep-th/0110244.  

\bibitem{GS} M. Gutperle and A. Strominger, {\it Timelike Boundary Liouville Theory}, 
Phys. Rev. D67 (2003) 126002; hep-th/0301038.

\bibitem{GM} P. Ginsparg and G. Moore {\it Lectures on 2D gravity and 2D string theory (TASI 1992)}, 
in {\it Recent directions in particle theory}, eds. J. Harvey and J. Polchinski, Proceedings of the 
1992 TASI, World Scientific, Sigapore (1993); hep-th/9304011.

\bibitem{Seiberg} N. Seiberg, {\it Notes on Quantum Liouville Theory and Quantum Gravity}, 
Progr. of Theor. Phys. Suppl. No. 102, 1990, p. 319. 

\bibitem{morecomments}  D. Kutasov and N. Seiberg, {\it More Comments on String Theory on $AdS_3$}, 
JHEP {\bf 9904} (1999) 008;  hep-th/9903219. 

\bibitem{Kostov}  I. K. Kostov, {\it Boundary Ground Ring in 2D String Theory}, 
Nucl. Phys. B689 (2004) 3-36; hep-th/0312301.

\bibitem{DFrKut}  P. Di Francesco and D. Kutasov, {\it 
 World Sheet and Space Time Physics in Two Dimensional (Super) String Theory}, 
 Nucl. Phys. B375 (1992) 119-172;  hep-th/9109005.
 
\bibitem{BD} N. D. Birrell and P. C. W. Davies, {\it Quantum Fields in Curved Space}, Cambridge 
Univ. Press, 1984.

\bibitem{Wald} R. M. Wald, {\it Quantum Field  Theory in Curved Spacetime and Black Hole Thermodynamics}, 
U. Chicago Press, 1994.
 
\bibitem{BT} N. D. Birrell and J. G. Taylor, {\it Analysis of interacting quantum field theory 
in curved spacetime}, J. Math. Phys. {\bf 21}(7), p. 1740, 1980. 
 
\bibitem{BF} N. D. Birrell and L. H. Ford, {\it Self-interacting Quantized Fields and Particle 
Creation in Robertson-Walker Universes}, Ann. of Phys. {\bf 122}, pp. 1-25 (1979). 

\bibitem{GR} L.S Gradshteyn and I.M. Ryzhik, {\it Tables of integrals, series, and products}, 
sixth edition, Academic Press. 

\bibitem{Ber} F. A. Berezin, {\it The method of second quantization}, Academic Press, 1966

\bibitem{FHL} D. Z. Freedman, M. Headrick and A. Lawrence, {\it On closed string tachyon dynamics},
 Phys. Rev. {\bf D 73} (2006) 066015; hep-th/0510126.

\bibitem{GuthPi} A.H.Guth and S.-Y. Pi, {\it Quantum mechanics of the scalar field in the new 
inflationary universe}, Phys. Rev. {\bf D 32}, p. 1899 (1985).

\bibitem{WW} E. J. Weinberg and A. Wu, {\it Understanding complex perturbative effective potentials}, 
Phys. Rev. {\bf D36}, p. 2474 (1987). 

\bibitem{DJ} S. R. Das and A. Jevicki, {\it String field theory and physical interpretation of 
D=1 strings}, Mod. Phys. Lett. {\bf A5}, p. 1639 (1990). 

\bibitem{LLM} N. Lambert, H. Liu and J. Maldacena, {\it Closed strings from decaying D-branes}, 
hep-th/0303139.

\bibitem{KMS} I. Klebanov, J. Maldacena and N. Seiberg, {\it  D-brane Decay in Two-Dimensional String Theory}, 
JHEP 0307 (2003) 045;  hep-th/0305159.


 \end{thebibliography}
\end{document}